\documentstyle[preprint,aps,psfig]{revtex}

\begin{document}

\title{$P$-matrix and $J$-matrix approaches. \\
Coulomb asymptotics in
the harmonic oscillator representation of scattering theory.}
\author{J.M.Bang$^1$, A.I.Mazur$^2$,
A.M.Shirokov$^{1,3,4}$\thanks{Permanent address: Skobeltsyn Institute
of Nuclear Physics, Moscow State University, Moscow
119899, Russia; E-mail: shirokov@anna19.npi.msu.su},
Yu.F.Smirnov$^{3,5}$, and S.A.Zaytsev$^{2,3}$% \\
\thanks{Permanent address:
Physics Department, Khabarovsk State Technical University,
Tikhookeanskaya 136, Khabarovsk 680035, Russia} }
\address{$^1$K{\o}benhavns Universitet, Niels Bohr Institutet, Blegdamsvej 17,
DK-2100 K{\o}benhavn \O, Denmark\\
$^2$Physics Department, Khabarovsk State Technical University,
Tikhookeanskaya 136, Khabarovsk 680035, Russia\\
$^3$Skobeltsyn Institute of Nuclear Physics, Moscow State University, Moscow
119899, Russia\\
$^4$International Institute of Theoretical and Applied Physics,
Iowa State University, Ames, Iowa 50011\\
$^5$Instituto de Ciencias Nucleares,
Universidad Nacional Aut\'{o}noma de M\'{e}xico,
M\'exico, D.F., 04510,
 M\'{e}xico}
%\date{ }
\maketitle
\begin{abstract}
The relation between the $R$- and $P$-matrix approaches and
the harmonic oscillator representation of the quantum scattering theory
($J$-matrix method) is discussed. We construct a
discrete analogue of the $P$-matrix  that is shown to
be equivalent to the usual $P$-matrix in the quasiclassical limit. A
definition of the natural channel radius is introduced. As a result,
it is shown to be possible to use
well-developed technique of $R$- and $P$-matrix theory for calculation
of resonant states characteristics, scattering phase shifts, etc., in the
approaches based on  harmonic oscillator expansions, e.g., in
nuclear shell-model calculations. $P$-matrix is used also for
formulation of the method of treating Coulomb asymptotics in the
scattering theory in oscillator representation.
\end{abstract}
\section{Introduction.}
\par
A number of methods for treating effects of the continuum spectrum in
nuclear structure calculations have been developed. We shall mention the 
$R$-matrix approach \cite{Wigner} (see also review paper \cite{Lane} and 
references
therein) and related to it the $P$-matrix approach (see \cite{Jaffe} and
references therein), the Feshbach projection method \cite{Fesh} and,
based on that, the continuum shell model \cite{Mahaux,Rotter}, early coupled
channel calculations \cite{Coupl,Coupl2,Coupl3,Coupl4} and more
developed various continuum RPA approaches \cite{Shlomo,RPA,Arima},
methods based on the expansions of the continuum spectrum wave function
in Sturmian (Weinberg) functions \cite{Sturm,Bang1} or in other basis
function sets \cite{Bang1}, the method of pole (Mittag-Leffler) expansion
\cite{Bang2,Bang1} of wave functions, Green function, scattering
amplitudes, etc.
\par
We feel, that it is very natural to use the so-called $J$-matrix
method in the studies of the continuum spectrum effects in nuclear
structure or in the low-energy nuclear scattering calculations. The
$J$-matrix method was initially proposed for atomic problems
\cite{Hell,YaFi} and  shown  to be one of
the most efficient and precise methods in calculations of
photoionization \cite{Broad,Stot} and electron scattering by atoms
\cite{Konov}. In nuclear physics the same approach has been developed
independently \cite{Fil,NeSm,Hung} as the method of harmonic oscillator
representation of scattering equations (HORSE). The HORSE method has been
successfully used in various  nuclear applications, e.g.,
nucleus-nucleus scattering  has been studied in the algebraic
version of RGM based on HORSE (see, e.g, the review papers
\cite{FVCh,Rev}); the effect of $\Lambda$ and neutron decay channels in
hypernuclei production reactions has been investigated in refs.
\cite{M1,M2}; the generalization of the HORSE method \cite{SmSh} to the
case of few-body channels within the democratic decay approximation has
been used in the cluster model calculations of monopole excitations in
$^{12}$C \cite{Mikh} and of the halo and soft dipole mode properties in
$^{11}$Li \cite{Lur} and $^6$He \cite{Lur-6He} (see also  review
papers~\cite{Viet,Cocoyoc}), in
the study of double-$\Lambda$ hypernuclei~\cite{doubleLambda}, etc.
\par
The HORSE method is
very attractive because the continuum spectrum wave function is
represented as a sum of oscillator functions, i.e., in the same manner
as in conventional variational nuclear structure calculations, say, in
shell model or cluster model RGM calculations. The wave functions,
$S$-matrix and other scattering characteristics are obtained by algebraic
methods that are
very efficient in calculations. As a result, it is possible
with the use of the HORSE method to treat  problems with a large number
of channels with moderate computer facilities (e.g., in $J$-matrix atomic
calculations \cite{Broad,Stot} up to $\sim 70$ channels have been
allowed for). With the HORSE method it is possible to perform the
calculation in the complex energy plane and to locate
$S$-matrix poles
numerically \cite{SmSh,OkhrS,Lur,Lur-6He,FilS}. The location of $S$-matrix
poles is important not only for calculation of resonance properties
\cite{Stot,OkhrS}, but also for bound states, e.g., for atomic systems
\cite{Stot} or for weakly-bound nuclear states \cite{Lur,Lur-6He} the location
of the $S$-matrix poles improves significantly the variational results
for binding energies. The expansion of the continuum spectrum wave
function in  oscillator function series involves, of course, an
infinite number of terms. However, usually only the first few terms of
the expansion are important in calculation of matrix elements of
operators of physical observables. Nevertheless, $N\hbar \omega$
terms with large values of $N$ can be easily calculated in the HORSE
method, and in some special cases these terms become
important, e.g., in calculation \cite{Lur} of electromagnetic transition
probabilities ${\cal B}(E\lambda; g.s.\to continuum)$ for weakly bound
systems terms up to $4000\hbar \omega$ have been allowed for.
\par
Below we shall discuss correlations between the HORSE method and the
well-known $R$-matrix and $P$-matrix approaches (the formal equivalence
between the $J$-matrix method and the Feshbach method \cite{Fesh} has been
proved in ref. \cite{Yamani}). We shall derive
exact expressions for $R$- and $P$-matrices within the HORSE formalism.
These expressions can be used for calculation of $R$- or $P$-matrices
within any variational approach based on the harmonic oscillator
expansion of the wave functions, e.g., in  shell-model calculations,
and thereafter the well-developed technique of $R$-matrix  and $P$-matrix
theory (see, e.g., \cite{Lane,Jaffe}) can be used for calculation of
scattering phase shifts, resonance energies and widths, etc., within the
variational approach. We shall show that within the HORSE formalism
a discrete analogue of the $P$-matrix can be introduced that  is
equivalent to the $P$-matrix in the quasiclassical limit.
\par
The construction of the $P$-matrix within the HORSE formalism gives the
possibility to formulate a very efficient method for introducing the
Coulomb asymptotics within the HORSE theory. The only method for
treating Coulomb interaction within the HORSE approach
has been developed by the Kiev group~\cite{Okhr}. The method of
ref.~\cite{Okhr} involves
diagonalization of large matrices even in a single-channel case that
makes it unusable for application to problems with a large number of channels.
Our method is free from these shortcomings and can be used for
high-accuracy calculations of scattering characteristics of
multichannel systems of charged particles.
\par
In this paper we discuss binary channels only.
\par
The paper is organized as follows. Basic equations of the HORSE approach
are presented in Section~2.  $P$-matrix and $R$-matrix are derived
within the HORSE formalism in  Section~3. In Section~4, we propose the
method of treating Coulomb interaction within the HORSE formalism
and demonstrate the accuracy of the proposed technique.
Section~5 is devoted to the generalization of the above
results to the multichannel case.
%Conclusions are formulated
Summary is presented in Section~6.
\section{Potential scattering in the HORSE method.}
\par
We  sketch in this section  the basic equations of the HORSE method.  We
restrict the discussion in this section to the simplest case of
potential single-channel scattering of uncharged particles.
\par
We use the conventional partial wave expansion of the wave  function,
 \begin{equation}
 \Psi^{\pm}_{\bf k}({\bf r}) = \frac{4\pi}{k}\sum_{l,m}
  i^{l} e^{\pm i\delta_l}
  u_{l}(k,r) Y^{*}_{lm}(\Omega_{\bf k})  Y_{lm}(\Omega_{\bf r}),
 \label{Ylm}
 \end{equation}
where ${\bf k}$ is momentum, $\Omega_{\bf k}$ and $\Omega_{\bf r}$ are
angular variables in momentum and coordinate spaces, respectively,
$Y_{lm}(\Omega)$ are the usual spherical functions, $\delta_{l}$ is the
phase shift in the partial wave labeled by the angular momentum $l$,
star is used to denote the complex conjugation. The partial amplitudes,
$u_{l}(k,r)$,
% are supposed to be
%normalized so that the flux associated with the wave function (\ref{Ylm})
%is equal to unity. They
are the eigenfunctions of the radial Schr\"{o}dinger equation,
 \begin{equation}
  H^l \,u_{l}(k,r)=E\,u_{l}(k,r)\ .
  \label{eq:Sh}
  \end{equation}
We normalize the partial amplitudes $u_{l}(k,r)$ in such a way that the
flux associated with the wave function (\ref{Ylm}) is equal to unity.
\par
Within the HORSE formalism, the partial amplitudes
$u_{l}(k,r)$ are expanded in  infinite series of the harmonic
oscillator functions,
  \begin{equation}
  u_{l}(k,r) = \sum_{n=0}^{\infty} a_{nl}(k) R_{nl}(r)\ .
  \label{eq:row}
  \end{equation}
In eq. (\ref{eq:row}),
  \begin{equation}
  R_{nl}(r) = (-1)^{n}\,\sqrt{ \frac{2\,n!}{r^{3}_{0} \Gamma (n+l+3/2)} }\,
  \left(\frac{r}{r_0}\right)^{l} \exp \left(-\frac{r^{2}}{2r_{0}^2}\right)
    \,L_{n}^{l+1/2}\left(\frac{r^{2}}{r_{0}^2}\right) \ ;
  \label{eq:oscrad}
  \end{equation}
$L_{n}^{\alpha}(x)$ is Laguerre polynomial; the oscillator
radius $r_{0}=\sqrt{\hbar / \mu \omega}$ is the only parameter of the
oscillator basis (\ref{eq:oscrad}); $\hbar \omega$ is the spacing
between the oscillator levels; $\mu$ is the reduced mass.
%; the energy
%$E\equiv \hbar \omega \varepsilon = \frac{\hbar \omega}{2}q^2$.
\par
The functions $a_{nl}(k)$ entering eq. (\ref{eq:row}) represent the wave
function in the oscillator representation. They obey the infinite set of
algebraic equations
  \begin{equation}
  \sum_{n'=0}^{\infty}\,(H_{nn'}^{l}-\delta_{nn'}E)\,
  a_{n'l}(k)=0,
  \label{eq:Infsys}
  \end{equation}
where the matrix elements of the Hamiltonian in the oscillator basis
$H_{nn'}^l =T_{nn'}^l + V_{nn'}^{l}$, and $T_{nn'}^l$ and $V_{nn'}^{l}$
are the matrix elements of kinetic and potential energy operators,
respectively. The matrix elements of short-range nuclear potentials
$V_{nn'}^{l} \to 0$ in the limit $n$ and/or $n' \to \infty$, while
the kinetic energy is represented by a tridiagonal matrix,
  \begin{equation}
   \begin{array}{l}
  T_{nn'}^l = 0\qquad \ \qquad \ $ if $\ \ \mid n-n' \mid\ > 1\ , \\
  T_{nn}^l =
    \displaystyle\frac{\displaystyle\hbar\omega}{\displaystyle 2}\,
                         (2n+l+3/2)\ , \\
  T_{n+1,n}^l =  T_{n,n+1}^l =
     \displaystyle  -\;\frac{\displaystyle\hbar\omega}{\displaystyle 2}
     \sqrt{(n+1)(n+l+3/2)}\ , \end{array}
  \label{eq:Tnm}
  \end{equation}
with non-zero matrix elements $T_{nn}^l$ and $T_{n,n\pm 1}^l$
increasing linearly with $n$ for large values of $n$. Thus, the
potential energy matrix can be truncated, and actually we shall use in
what follows instead of $V$ the potential energy $\tilde{V}$ with matrix
elements
%in the oscillator representation defined by the following equation:
  \begin{equation}
    \tilde{V}_{nn'}^l = \left\{
    \begin{array}{lll}
          \ V_{nn'}^l \ \ \ \ \ &$ if $ \ \ & n$ and $n'\leq N;\\
          \ 0 &$ if $& n $ and/or $ n'>N\ .
     \end{array} \right.
  \label{trunc}
  \end{equation}
The truncation (\ref{trunc}) is the only approximation of the HORSE
method. Note, that the kinetic energy matrix is not truncated within the
HORSE approach in contrast to the conventional oscillator-basis
variational methods which involve the diagonalization of the
truncated Hamiltonian matrix $\tilde{H}_{nn'}^l$. Thus, the HORSE method
is equivalent to the method of high-rank separable approximation of the
potential.
\par
We shall introduce in the oscillator representation the interaction
region spanned by the functions (\ref{eq:oscrad}) with $n\leq N$ and the
asymptotic region spanned by the functions (\ref{eq:oscrad}) with $n>N$.
In the asymptotic region, the wave function in oscillator
representation, $a_{nl}^{as}(k)$, obeys the three-term recurrence
relation
  \begin{equation}
  T_{n,\;n-1}^l \,a_{n-1,\;l}^{as}(k) +
  (T_{nn}^l - E)\,a_{nl}^{as}(k) +
   T_{n,\;n+1}^l \,a_{n+1,\;l}^{as}(k) = 0\ ,
  \label{eq:TRS}
  \end{equation}
as is easily seen from eqs. (\ref{eq:Infsys}), (\ref{eq:Tnm}) and
(\ref{trunc}). Eq. (\ref{eq:TRS}) has two linearly independent
solutions, $S_{nl}(k)$ and $C_{nl}(k)$,
and $a_{nl}^{as}(k)$ can be expressed as a linear
combination of these solutions:
  \begin{equation}
  a_{nl}^{as}(k)=\cos \delta_{l}\;S_{nl}(k)+
                                  \sin \delta_{l}\;C_{nl}(k), \,\,\,\,
  n \geq N.
  \label{eq:asosc}
  \end{equation}
It is convenient to use the following set of linearly independent
solutions $S_{nl}(k)$ and $C_{nl}(k)$ \cite{YaFi,NeSm}:
%  \begin{equation}
\begin{eqnarray}
  S_{nl}(k) = & & \sqrt{\frac{\pi\, r_{0}\, n!}{v \Gamma (n+l+3/2)}}\,
    (kr_{0})^{l+1}\, \exp \left(-\frac{k^2 r_{0}^{2}}{2} \right)\,
   L^{l+1/2}_{n}(k^{2}r^{2}_0)\ ,
  \label{eq:Snl}     \\
%  \end{equation}
%  \begin{eqnarray}
  C_{nl}(k) = & &\frac{(-1)^l}{\Gamma (-l+1/2)} \,
  \sqrt{ \frac{\pi\, r_{0}\,n!}{v \Gamma (n+l+3/2)} }\,(kr_0 )^{-l}\,
    \exp \left(-\frac{k^2 r_{0}^{2}}{2} \right) \nonumber \\
  & & \times \Phi (-n-l-1/2,\, -l+1/2;\, k^2 r_{0}^{2})\ ,
  \label{eq:Cnl}
  \end{eqnarray}
where the velocity $v=\hbar k/\mu$, and
$\Phi (a,\, b;\, z)$ is a confluent hypergeometric function
\cite{Erd1}. The solutions (\ref{eq:Snl}) and (\ref{eq:Cnl}) are defined
in such a way that
\begin{equation}
 \sum_{n=0}^{\infty} S_{nl}(k)R_{nl}(r) = \frac{k}{\sqrt{v}}\; j_{l}(kr)\ ,
  \label{S-j}
\end{equation}
and in the limit $r\to\infty$
\begin{equation}
\sum_{n=0}^{\infty} C_{nl}(k)R_{nl}(r) \to -\frac{k}{\sqrt{v}}\; n_{l}(kr)\
. \label{C-n}
\end{equation}
Here  $j_{l}(x)$ and $n_{l}(x)$ are spherical Bessel and Neumann functions
\cite{Erd2,Abr}, respectively. Thus, the adopted
normalization of the wave function (\ref{Ylm}) is assured, and
%corresponding to the unit flux is assured, and
$\delta_{l}$ entering eq. (\ref{eq:asosc}) appears to be just
the phase shift in the partial wave corresponding to the angular
momentum~$l$.
%\cite{YaFi,NeSm}.
\par
The functions (\ref{eq:Snl}) and  (\ref{eq:Cnl}) can be
easily calculated. The
% confluent hypergeometric function entering eq.
%(\ref{eq:Snl}) can be reduced to the Laguerre polynomial
%$L_{n}^{l+1/2}(k^{2} r_{0}^2 )$ (see, e.g., \cite{Erd2}), thus the
function $S_{nl}(k)$ is seen  to be the
harmonic oscillator function in the momentum representation.
\par
We shall need in what follows  asymptotics  of the functions
(\ref{eq:Snl}) and (\ref{eq:Cnl}) in the limit $n\to \infty$:
 \newcounter{abc}
 \addtocounter{equation}{1}
 \setcounter{abc}{\value{equation}}
 \setcounter{equation}{0}
 \renewcommand{\theequation}{\arabic{abc}\alph{equation}}
\begin{eqnarray}
  S_{nl}(k) & \approx &2kr_0\sqrt{\frac{r_{0}}{v} }\,
   (n+l/2+3/4)^{\frac{1}{4}}\,
         j_{l}(2kr_{0}\sqrt{n+l/2+3/4})
  \label{Snlass}  \\
& \approx &\sqrt{\frac{r_{0}}{v} }\, (n+l/2+3/4)^{-\frac{1}{4}}\,
         \sin[2kr_{0}\,\sqrt{n+l/2+3/4}-\pi l/2]\ ,
  \label{eq:Snlass}
\end{eqnarray}
and
  \addtocounter{abc}{1}\setcounter{equation}{0}
\begin{eqnarray}
  C_{nl}(k) & \approx &-2kr_0\sqrt{\frac{r_{0}}{v} }\,
   (n+l/2+3/4)^{\frac{1}{4}}\,
         n_{l}(2kr_{0}\sqrt{n+l/2+3/4})
  \label{Cnlass}  \\
& \approx &\sqrt{\frac{r_{0}}{v} }\, (n+l/2+3/4)^{-\frac{1}{4}}\,
         \cos[2kr_{0}\,\sqrt{n+l/2+3/4}-\pi l/2]\ ,
  \label{eq:Cnlass}
\end{eqnarray}
that can be easily obtained from the asymptotics of the confluent
hypergeometric functions \cite{Erd1} and Laguerre polynomials \cite{Erd2}.
  \renewcommand{\theequation}{\arabic{equation}}
  \setcounter{equation}{\value{abc}}
\par
The so-called Casoratian determinant, ${\cal K}^{l}_{n}(C,S) \equiv
C_{n+1,\,l}(k)\,S_{nl}(k)\,-\,C_{nl}(k)\,S_{n+1,\,l}(k)$, plays the same
role in the theory of second-order finite difference equations like
eq.~(\ref{eq:TRS}) (see, e.g., \cite{MirSol,Korn}) as Wronskian in the theory
of second-order differential equations. It is easy to show that
${\cal A} \equiv T_{n,\,n+1}^{l} {\cal K}^{l}_{n}(C,S)$ is $n$-independent.
Thus, ${\cal A}$ can be easily calculated in the limit $n\to\infty$
using asymptotics (\ref{eq:Snlass}) and (\ref{eq:Cnlass}). As a result, we
get:
\begin{equation}
    T_{n,\,n+1}^{l} {\cal K}^{l}_{n}(C,S) = \frac{\hbar}{2}\ .
  \label{A}
  \end{equation}
\par
In the interaction region $n \leq N$, the set of algebraic equations
(\ref{eq:Infsys}) due to (\ref{eq:Tnm}) and (\ref{trunc}) takes the
following form:
  \begin{equation}
  \sum_{n'=0}^{N}\,(H^{l}_{nn'}-
   \delta_{nn'}E )\,a_{n'l}(k) =
  -\delta_{nN}\;T^{l}_{N,\, N+1}\,a_{N+1,\, l}^{as}(k);
        \,\,\,\, n=0,1,\ldots ,N\ .
  \label{eq:OsSh}
  \end{equation}
The functions $a_{nl}(k)$ satisfying eqs. (\ref{eq:OsSh}) in the
interaction region $n \leq N$ can be expressed in terms of
$a_{N+1,\,l}^{as}(k)$,
  \begin{equation}
  a_{nl}(k) = {\cal G}_{nN}\,a_{N+1,\, l}^{as}(k)\ ,
  \label{eq:cnin}
  \end{equation}
and  the matrix elements ${\cal G}_{nn'}$ can be calculated by the
formula
  \begin{equation}
  {\cal G}_{nn'} = -\sum_{\lambda=0}^{N}\,
  \frac{ \gamma_{\lambda n}^{*}\,\gamma_{\lambda n'} }
  { E_{\lambda}\,-\,E }\,T^{l}_{n'\!\!,\; n'+1}\ .
  \label{eq:oscrm}
  \end{equation}
Here $E_{\lambda }$ and $\gamma_{\lambda n}$ are eigenvalues and
corresponding eigenvectors of the truncated Hamiltonian matrix
$\tilde{H}_{nn'}^{l}$, $n,n'=0,1,\ldots ,N$.
\par
The function  $a_{Nl}(k)$ should fit both eqs. (\ref{eq:TRS})
and (\ref{eq:OsSh}). Thus, we can express both $ a_{Nl}(k)$ and
$a_{N+1,\, l}^{as}(k)$ in terms of the phase shift $\delta_{l}$ using eq.
(\ref{eq:asosc}), and insert these expressions in eq. (\ref{eq:cnin}).
As a result, we obtain a simple formula \cite{YaFi} for %the calculation of
the phase shift $\delta_{l}$:
  \begin{equation}
  \tan \delta_{l}\,=\,-\frac
  { S_{Nl}(k)\,-\,{\cal G}_{NN}S_{N+1,\, l}(k) }
    { C_{Nl}(k)\,-\,{\cal G}_{NN}C_{N+1,\, l}(k) }\ .
  \label{eq:oscph}
  \end{equation}
\par
So, within the HORSE method, one should first diagonalize
the truncated Hamiltonian matrix
$\tilde{H}_{nn'}^{l}$, $n,n'=0,1,\ldots ,N$, i.e. one should find
eigenvalues $E_{\lambda}$ and eigenvectors $\gamma_{\lambda n}$.
Next, the  matrix ${\cal G}_{nn'}$ and the phase shift $\delta_{l}$
are calculated using eqs. (\ref{eq:oscrm}) and (\ref{eq:oscph}) while
$S_{nl}(k)$ and $C_{nl}(k)$ entering eq. (\ref{eq:oscph}) are
calculated using (\ref{eq:Snl}) and (\ref{eq:Cnl}). The functions
$a_{nl}(k)$ defining the wave function in the oscillator representation,
are to be obtained by eq. (\ref{eq:asosc}) in the asymptotic
region $n \geq N$ and by eq. (\ref{eq:cnin}) in the interaction region
$n \leq N$.
\par
All the above expressions can be generalized to the case of
multichannel scattering \cite{YaFi} or to the case of true few-body
scattering (democratic decay approximation) \cite{SmSh}.
The convergence of all of the physical observables in nuclear
calculations is usually obtained if the interaction region for each
channel is spanned by 5--10 oscillator functions (see, e.g.,
\cite{Hung,FVCh,Rev,M1,M2,Mikh,Lur}).
\par
The truncated Hamiltonian matrix
$\tilde{H}_{nn'}$ is actually the same matrix that is used in conventional
variational approaches with oscillator basis, e.g., in the shell-model
calculations. The HORSE method is a natural generalization of these
variational approaches on the case of continuum spectrum states.
The calculation of the matrix
elements of the truncated Hamiltonian matrix $\tilde{H}_{nn'}$
and its diagonalization is
the most complicated part of the calculations within the HORSE approach.
An important point concerning the
applications of the HORSE method is that the diagonalization of
$\tilde{H}_{nn'}$ should be done only once, thereafter
the negative eigenvalues
$E_{\lambda}$ can be interpreted as the energies of the bound states and
the corresponding eigenvectors $\gamma_{\lambda n}$ can be used
instead of $a_{nl} $ in the expansion (\ref{eq:row})
for the construction of the bound state wave functions, while continuum
spectrum wave functions in a wide range of positive values of energy $E$ 
are easily calculated by
means of simple formulas (\ref{eq:asosc}), (\ref{eq:Snl}),
(\ref{eq:Cnl}), (\ref{eq:cnin}), (\ref{eq:oscrm}) and (\ref{eq:oscph}).
\section{Discrete analogue of the $P$-matrix and natural channel radius.}
\par
The $P$-matrix is identical to the inverse $R$-matrix and is defined
\cite{Lane,Jaffe} as the
logarithmic derivative of the wave function at the channel radius $r=b$:
 \begin{equation}
   P = R^{-1} = \frac{\left. b\,\frac{d}{dr}u_{l}(k,r)
   \right|_{r=b}}{u_{l}(k,b)} \ .
  \label{eq:Popr}
  \end{equation}
The partial amplitudes ${\bar u}_{l}(k,r) = ru_{l}(k,r)$ are often used in
applications. We can define also $P$-matrix $\bar P$ by the expression
(\ref{eq:Popr}) with ${\bar u}_{l}(k,r)$ instead of $u_{l}(k,r)$. It is easy
to see that
\begin{equation}
    P = {\bar P} - 1 \,.
         \label{barP}
\end{equation}
\par
The channel radius $b$ is a parameter of the $P$- and $R$-matrix
theories;  $b$ divides the whole coordinate space into two parts:
the ``interaction
region" $r\le b$ and the ``asymptotic region" $r\ge b$, i.e. the channel radius
$b$ plays the same role as the truncation boundary $N$ in the
oscillator space of the HORSE method.
In the asymptotic region $r\ge b$, the wave function is of the form:
  \begin{equation}
  u_{l}(k,r)= \cos \delta_{l}\,{\hat F}_{l}(k,r)+
                                  \sin \delta_{l}\,{\hat G}_{l}(k,r)\ .
  \label{eq:assN1}
  \end{equation}
\par
If some long-range
interaction, say, Coulomb interaction, cannot be neglected in the
asymptotic region $r\ge b$, than ${\hat F}_{l}(k,r)$ and ${\hat G}_{l}(k,r)$
are respectively the regular and irregular solutions of the radial
Schr\"{o}dinger equation involving the corresponding potential.
In the simplest case, when the interaction between particles can be
neglected in the asymptotic region $r\ge b$,
${\hat F}_{l}(k,r) =
\frac{\displaystyle k}{\displaystyle\sqrt{v}}\, {j}_{l}(kr)$ and
${\hat G}_{l}(k,r) =
-\frac{\displaystyle k}{\displaystyle \sqrt{v}}\,{n}_{l}(kr)$.
%where ${j}_{l}(kr)$  and ${n}_{l}(kr)$ are the spherical Bessel and
%Neumann functions \cite{Erd2,Abr}, respectively.
If, additionally, the
channel radius $b$ is large enough and spherical Bessel and
Neumann functions
%${j}_{l}(kr)$  and ${n}_{l}(kr)$
can be replaced by their  asymptotics at $r\geq b$,
then in the asymptotic region
  \begin{equation}
  u_{l}(k,r)=\frac{1}{r\sqrt{v}}\,
%[\cos \delta_{l}\sin (kr-\pi l/2)+ \sin \delta_{l}\cos (kr-\pi l/2)]\ .
     \sin \Bigl(kr - \frac{\pi l}{2} + \delta_{l}\Bigr)\ .
  \label{eq:assN}
  \end{equation}
In this case, from eq. (\ref{eq:Popr}) it follows, that
  \begin{equation}
  P = R^{-1} =
%kb\,\frac{1-\tan(kb-\pi l/2)\,\tan \delta_{l}}
    kb\, \cot\Bigl(kb - \frac{\pi l}{2} + \delta_{l}\Bigr)\ ,
  \label{eq:Pmat}
  \end{equation}
while in the general case, we have
  \begin{equation}
  P = R^{-1} = b\,\frac{{\hat F}_{l}^{\prime}(k,r)
      + \tan \delta_{l} {\hat G}_{l}^{\prime}(k,r) }
    {{\hat F}_{l}(k,b)  + \tan \delta_{l} {\hat G}_{l}(k,b) }\ ,
  \label{genP}
  \end{equation}
where ${\hat F}_{l}^{\prime}(k,b)\equiv \left.
\frac{\displaystyle d}{\displaystyle dr}{\hat F}_{l}(k,r)\right|_{r=b}$ and
${\hat G}_{l}^{\prime}(k,b) \equiv \left.
\frac{\displaystyle d}{\displaystyle dr}{\hat G}_{l}(k,r)\right|_{r=b}\ $.
\par
According to the general
theory of ref. \cite{Lane}, in the interaction region $r\le b$
the wave function $u_{l}(k,r)$ is expanded in a complete set of
functions $\left\{v_{nl}(r)\right\}$
matching some boundary condition at $r=b$, e.g., in the $P$-matrix
approach $v_{nl}(r)$ satisfy the boundary condition $v_{nl}(b)=0$. Using
the formalism described in detail in refs. \cite{Lane,Jaffe}, $P$- or
$R$-matrix can be calculated solving the Schr\"{o}dinger equation in the
functional space spanned by the functions $\left\{v_{nl}(r)\right\}$,
and thereafter the phase shift $\delta _{l}$ can be found using eq.
(\ref{eq:Pmat}) or (\ref{genP}). The well-developed formalism of $R$- or
$P$-matrix theory can be also used for calculations of resonance
positions and widths.
\par
It is seen that the
HORSE formalism has much in common with the $R$-matrix and $P$-matrix
ones.
% as it has been already
%mentioned in the first paper devoted to this approach \cite{Hell}.
All the above mentioned formalisms involve the division
of the whole space into the interaction and asymptotic regions.
Nevertheless, within the HORSE approach one divides the
functional space, while within the $R$- and $P$-matrix approaches the
coordinate space is divided into two regions.
The HORSE formalism involves the oscillator basis that
differs significantly from the basises used in the $R$-matrix or in the
$P$-matrix theory, and there is no formal equivalence between the
HORSE method and the $R$-matrix method or the $P$-matrix method. The
HORSE formalism is very attractive because oscillator basis
is natural in nuclear structure studies, while $R$- and
$P$-matrix formalisms involve basises that are more artificial for nuclear
physics applications.
\par
Applications of the $P$-matrix formalism \cite{Jaffe,Sim} refer to
a very important property of the $P$-matrix: the
$P$-matrix has poles at the energies that can be identified with the
energies of the so-called `primitives'~\cite{Jaffe}, i.e., of the eigenstates
obtained with an artificial boundary condition $v_{nl}(b)=0$. This feature
correlates scattering phase shifts, resonance positions  and widths with
the structure of the spectrum of the Hamiltonian in the interaction
region in the $P$-matrix approach. The boundary condition of the type
$v_{nl}(b)=0$ is good for quark bag models \cite{Jaffe,Sim}, but it is
dubious for nuclear structure applications.
If one starts from the nuclear shell model, than it is not clear what value
should be attributed to the channel radius $b$ and what is the correlation
between the primitives and the shell
model states. Thus, it is interesting to derive the $P$-matrix within
the HORSE formalism.
\par
Let us suppose that we can neglect the interaction at distances
$r \ge b$. Then for calculation of the $P$- or $R$-matrix one can use eq.
(\ref{genP}) or eq. (\ref{eq:Pmat}). Substituting
$\tan \delta_l$ in eq. (\ref{genP}) by the expression
(\ref{eq:oscph}), we obtain the following expression for $P$- and
$R$-matrices within the HORSE formalism in the general case:
  \begin{equation}
  P = R^{-1} = b\,\frac{ C_{Nl}(k) j_{l}^{\prime}(kb) +
     S_{Nl}(k)n_{l}^{\prime}(kb) -
  {\cal G}_{NN} \left[ C_{N+1,\, l}(k) j_{l}^{\prime}(kb) +
    S_{N+1,\, l}(k) n_{l}^{\prime}(kb) \right]  }
       { C_{Nl}(k) j_{l}(kb) +
     S_{Nl}(k) n_{l}(kb)  -
  {\cal G}_{NN} \left[ C_{N+1,\, l}(k) j_{l}(kb) +
    S_{N+1,\, l}(k) n_{l}(kb)  \right]  }\ ,
  \label{Pgen}
  \end{equation}
where
%${f}_{l}(kr) = (kr)\,{j}_{l}(kr)$,
%${g}_{l}(kr) = -(kr)\,{n}_{l}(kr)$,
$j_{l}^{\prime}(kb) \equiv \left.
\frac{\displaystyle d}{\displaystyle dr}j_{l}(kr)\right|_{r=b}$ and
$n_{l}^{\prime}(kb) \equiv \left.
\frac{\displaystyle d}{\displaystyle dr}n_{l}(kr) \right|_{r=b}\;$. In the
case when the approximation (\ref{eq:assN}) can be used, expression
(\ref{Pgen}) can be simplified and it takes the following form:

  \begin{eqnarray}
   & &\lefteqn{ P  = R^{-1}  }
           \nonumber   \\*
   & & \ \, = kb\,\frac{C_{Nl}(k)+
   \tan \left( kb- \frac{\pi l}{2}\right) S_{Nl}(k) -
      {\cal G}_{NN} \left[ C_{N+1,\,l}(k)+
     \tan \left( kb-\frac{\pi l}{2}\right) S_{N+1,\,l}(k)\right] }
    {C_{Nl}(k) \tan \left( kb-\frac{\pi l}{2}\right) - S_{Nl}(k) -
      {\cal G}_{NN}\left[ C_{N+1,\,l}(k)
 \tan \left( kb-\frac{\pi l}{2}\right) -S_{N+1,\,l}(k)\right] } - 1.
  \label{eq:Posc}
  \end{eqnarray}
\par
Eqs. (\ref{Pgen}) and (\ref{eq:Posc}) are very general and can be used
within the HORSE formalism for any value of the channel radius $b$ large
enough to neglect the potential energy at  $r\geq b$. These
equations can be used for calculation of $P$- or $R$-matrix within the
shell model approach with the subsequent use of the well-developed $R$-
or $P$-matrix formalism for calculations of phase shifts, properties of
resonances, etc.
\par
It is seen, that the $P$-matrix poles do not coincide with the
eigenenergies of the truncated Hamiltonian matrix that are actually
the poles of ${\cal G}_{NN}$ as  evident from the definition
(\ref{eq:oscrm}). Thus, generally the oscillator shell model states
cannot be associated with the poles of the $P$-matrix. To put the $P$-matrix
poles in the one-to-one correspondence with the shell model states, we
should remove ${\cal G}_{NN}$ from the denominator of eq. (\ref{Pgen}).
This can be done if the channel radius $b$  fits the following
equation:
   \begin{equation}
   \frac{j_{l}(kb)}{n_{l}(kb)} =
     -\; \frac{   S_{N+1,\, l}(k) }{C_{N+1,\, l}(k) }\ .
  \label{frac}
   \end{equation}
\par
Eq. (\ref{frac}) has an infinite number of energy-dependent solutions
$b_i$. Nevertheless, in the quasiclassical limit $N\to \infty$, one of the
solutions  becomes energy-independent. This solution, $b_0$, is easily
found substituting
$S_{N+1,\, l}(k) $ and $C_{N+1,\, l}(k)$ in eq. (\ref{frac}) by
their asymptotics (\ref{Snlass}) and  (\ref{Cnlass}):
  \begin{equation}
  b_0 = 2r_{0}\,\sqrt{N+\frac{l}{2}+\frac{7}{4}}\ .
  \label{eq:a}
  \end{equation}
\par
Eq.~(\ref{frac}) may be reduced to a much simpler form if $j_{l}(kb)$ and
$n_{l}(kb)$ are replaced by their asymptotics that is equivalent to the use
of the approximation (\ref{eq:assN}) for the wave function in the asymptotic
region and eq.~(\ref{eq:Posc}) for $P$-matrix:
\begin{equation}
  \tan \Bigl(kb-\frac{\pi l}{2}\Bigr)\,=\, \frac { S_{N+1,\, l}(k) }
{ C_{N+1,\, l}(k) }\ .
     \label{eq:usl1}
  \end{equation}
This equation can be easily  solved, and we obtain:
  \begin{equation}
  b_{i} = \frac{\pi l}{2k}\, +
   \frac{1}{k}\arctan\left( \frac{ S_{N+1,\, l}(k)}{C_{N+1,\, l}(k)}\right)
   + \frac{i \pi}{k}\ ,
       \,\,\,\,\, i=0,\pm 1,\pm 2, \ldots\ \  .
  \label{eq:ka}
  \end{equation}
The solution (\ref{eq:a}) corresponds to $i=0$ that is easily seen
substituting
$S_{N+1,\, l}(k) $ and $C_{N+1,\, l}(k)$ in eq.~(\ref{eq:ka}) by
their asymptotics (\ref{eq:Snlass}) and  (\ref{eq:Cnlass}). It is seen that
all the rest solutions $b_{i} \to b_{0}$ in the limit $k \to \infty$.
\par
We shall refer to the energy-independent channel radius defined
according to eq.~(\ref{eq:a}) as a {\em natural channel radius} $b_0$.
The  natural channel radius $b_0$ coincides with the classical turning
point, $r_{N+1}^{cl}=2r_{0}\,\sqrt{N+l/2+7/4}$, corresponding to  the
oscillator function $R_{N+1,\,l}(r)$, i.e.\ to the first function
outside the interaction region in the oscillator representation used for the
%that is characterized
%by the maximum value of $r_{n}^{cl}$ among all the functions used for the
construction of the truncated Hamiltonian matrix $\tilde{H}_{nn'}$.
\par
It is worthwhile to note, that though  eq.~(\ref{eq:a}) has been obtained
using the asymptotics (\ref{Snlass}) and  (\ref{Cnlass}), it fits well
one of the solutions of eq.~(\ref{frac}) even for very small values of
$N$. This is illustrated by fig.~1 where the relative deviation,
$\displaystyle \frac{b_0-b}{b}$, of the natural channel radius $b_0$ in
various partial waves  from the exact solution $b$ of eq.~(\ref{frac}) is
plotted vs energy $E$.  The calculations are performed with
$\hbar\omega =18$~MeV and reduced mass corresponding to the neutron
scattering by $A=15$ nucleus. The absolute value of
$\displaystyle\frac{b_0-b}{b}$ is seen from the figure to increase
linearly with energy. However, even in the limiting case $N=0$ (only
one oscillator function present in the interaction region), the relative
deviation $\displaystyle \frac{b_0-b}{b}$ does not exceed 8\% only.
\par
 Using eqs. (\ref{A})
and (\ref{Pgen}) we obtain for the $P$-matrix
corresponding to the channel radius $b_i$ matching eq. (\ref{frac}) the
following expression:
  \begin{eqnarray}
\lefteqn{
  P = \frac{2b_{i} T_{N,\,N+1}^{l}}{\hbar}\,\tilde{n}_{l}(kb_{i})\,
   C_{N+1,\,l}(k) } \nonumber \\
 & &\ \ \ \ \ \ \ \times \, \left\{
    \breve{\jmath}_{l}(kb_{i})\,C_{Nl}(k)\,+\,
   S_{Nl}(k)\,-\,{\cal G}_{NN}\,\left[ \breve{\jmath}_{l}(kb_{i})\,
   C_{N+1,\,l}(k)\,+\,S_{N+1,\,l}(k)\right] \right\}\ ,
  \label{Pi}
  \end{eqnarray}
where $\breve{\jmath}_{l}(kb) \equiv j_{l}^{\prime}(kb)/n_{l}^{\prime}(kb)$
and $\tilde{n}_{l}(kb) \equiv n_{l}^{\prime}(kb)/n_{l}(kb)$.
In the approximation (\ref{eq:assN}), eq.~(\ref{Pi}) reduces to
  \begin{equation}
  P = -\frac{2kb_{i} T^{l}_{N,\, N+1}}{\hbar}\!
     \left\{ C_{N+1,\,l}(k)C_{Nl}(k)\!+\!S_{N+1,\,l}(k)S_{Nl}(k) -
      {\cal G}_{NN}\! \left[ C_{N+1,\,l}^{2}(k)+
      S_{N+1,\,l}^{2}(k)\right]\! \right\}\!-\!1.
  \label{eq:Posc1}
  \end{equation}
\par
Further simplifications of eq. (\ref{eq:Posc1}) can be introduced using
the asymptotics (\ref{eq:Snlass}) and  (\ref{eq:Cnlass}) of the functions
$C_{nl}(k)$ and $S_{nl}(k)$. After some straightforward algebra, in the
most interesting case $b=b_0$ we derive a very simple expression
connecting $P$-matrix and ${\cal G}_{NN}$:
  \begin{equation}
 P \ = \  2\;\sqrt{\Bigl(N+1\Bigr)\: \Bigl(N+l+\frac{3}{2}\Bigr)}\
     \Bigl(  \beta  - {\cal G}_{NN}  \Bigr)\ -\ 1\ ,
   \label{Pdis3}
  \end{equation}
where
  \begin{equation}
  \beta =
    \left(\frac{2N+l+7/2}{2N +l+3/2}\right)^{\frac{1}{4}} \cos \left[
      2kr_{0} \left( \sqrt{N+\frac{l}{2}+\frac{7}{4}} -
     \sqrt{N+\frac{l}{2}+\frac{3}{4}}\;\right)\right] .
          \label{beta}
\end{equation}
Obviously, $\beta\to 1$ in the limit of large $N$, and
$\sqrt{(N+1)(N+l+3/2)} = N+l/2+5/4 + O\bigl(N^{-1}\bigr)$. Thus, we can
rewrite eq.~(\ref{Pdis3}) as
 \begin{equation}
 P = 2\Bigl(N+\frac{l}{2}+\frac{5}{4}\Bigr)\,\Bigl(1-{\cal G}_{NN}\Bigr)
         - 1\ ,
  \label{eq:Pdis3}
  \end{equation}
or as
  \begin{equation}
   P \ = \ \left(2N+l+\frac{5}{2}\right)\,
   \frac{a_{N+1,\,l} (k)-a_{Nl} (k)}{a_{N+1,\,l} (k)}\ - \ 1
  \label{eq:Pdiskr}
  \end{equation}
using eq.~(\ref{eq:cnin}).
\par Equation (\ref{eq:Pdiskr}) obtained in the quasiclassical limit
$N\to\infty$ can serve, however, as a good approximation for the exact
$P$-matrix even for very small values of $N$. As is seen from fig.~2a,
even in the case $N=1$ the $P$-matrix calculated using
eq.~(\ref{eq:Pdiskr}) is very close to the exact one calculated by
eq.~(\ref{Pgen}).  However, there is a small discrepancy between the
positions of the poles of the $P$-matrices and, as a result, between
their values in the vicinity of the poles. As the number of oscillator
functions in the interaction region increases, the discrepancy
disappears and in the case $N=9$ the low-energy poles coincide (see
fig.~2b). However, there is still a small discrepancy between the poles
at higher energies that arises from the small increase with energy of
the channel radius $b$ fitting eq.~(\ref{frac}) while the natural
channel radius $b_0$ is energy-independent (see fig.~1).
\par
It is interesting to compare eq.~(\ref{eq:Pdiskr}) with
eqs.~(\ref{eq:Popr})--(\ref{barP}).
The first term in eq.~(\ref{eq:Pdiskr}) looks like a finite-difference
analogue of the $P$-matrix ${\bar P}$ with the wave
function in oscillator representation $a_{N+1,\,l} (k)$ and
finite difference $\Delta a_{N+1,\,l} (k) \equiv
a_{N+1,\,l} (k)-a_{Nl} (k)$ playing the roles of
the wave function ${\bar u}_{l}(k,b)$ and its derivative
$\left. \frac{d}{dr}{\bar u}_{l}(k,r) \right|_{r=b}$, respectively.
\par
To make this analogy more transparent, we note that the range of an
oscillator function $R_{nl}(r)$ is characterized by its classical turning
point, $r^{cl}_n = 2r_{0}\sqrt{n+l/2+3/4}$. Thus, in the definition of the
discrete analogue of the $P$-matrix we should use instead of
$\left. \frac{d}{dr}{\bar u}_{l}(k,r) \right|_{r=b}$ its finite-difference
analogue,
\begin{equation}
\frac{\Delta a_{N+1,\,l} (k)}{\Delta r^{cl}_{N+1}}\: \equiv \:
\frac{a_{N+1,\,l} (k)-a_{Nl} (k)}{r^{cl}_{N+1}-r^{cl}_{N}}\: =\:
\Bigl[ a_{N+1,\,l} (k)-a_{Nl} (k)\Bigr] \frac{1}{r_0}\
\sqrt{N+\frac{l}{2}+\frac{5}{4}}\ \Bigl[ 1+O\Bigl( N^{-1}\Bigr)\Bigr] .
\label{fdif}
\end{equation}
Contrary to the coordinate space, we have two characteristic lengths in the
harmonic oscillator space instead of a single length parameter $b$: the
range of the interaction space, $r^{cl}_N$, equal to the range of the
last basis function $R_{Nl}(r)$ included in the interaction region, and the
range of the first basis function $R_{N+1,\,l}(r)$ included in the
asymptotic region, $r^{cl}_{N+1}$. Defining the discrete analogue of the
$P$-matrix, ${\bar{\cal P}}_N$, we replace $b$ by a symmetrical combination
of both lengths, $\sqrt{r^{cl}_{N}\: r^{cl}_{N+1}}=
2r_{0}[(N+l/2+3/4)(N+l/2+7/4)]^{1/4}=2r_{0}\sqrt{N+l/2+5/4}\;[1+O(N^{-1})]$.
As a result, for the discrete analogue of the $P$-matrix ${\bar P}$ we
obtain the following expression:
  \addtocounter{equation}{1}
 \setcounter{abc}{\value{equation}}
 \setcounter{equation}{0}
 \renewcommand{\theequation}{\arabic{abc}\alph{equation}}
\begin{eqnarray}
  {\bar{\cal P}}_N & \equiv & \sqrt{r^{cl}_{N}\: r^{cl}_{N+1}}\
    \frac{1}{a_{N+1,\,l} (k)}\;
       \frac{\Delta a_{N+1,\,l} (k)}{\Delta r^{cl}_{N+1}}
         \label{disaP-}   \\
    & = & \left(2N+l+\frac{5}{2}\right)\,
   \frac{a_{N+1,\,l} (k)-a_{Nl} (k)}{a_{N+1,\,l} (k)}\
           \label{disaP-=}
\end{eqnarray}
(terms proportional to $N^{-1}$ and
higher-order corrections are omitted in eq.~(\ref{disaP-=})). So, we have
shown that the first term in eq.~(\ref{eq:Pdiskr}) is really the discrete
analogue ${\bar{\cal P}}_N$ of the $P$-matrix ${\bar{P}}$.
   \renewcommand{\theequation}{\arabic{equation}}
  \setcounter{equation}{\value{abc}}
\par
The $P$-matrix ${\bar P}$ involves the partial amplitudes
${\bar u}_{l}(k,r)$ while the $P$-matrix ${P}$ involves the partial
amplitudes ${u}_{l}(k,r) = \frac{1}{r} {\bar u}_{l}(k,r)$. By analogy,
in the harmonic oscillator representation we define functions
$\alpha_{nl}(k) = \frac{1}{r^{cl}_{n}} a_{nl}(k)$. Replacing $a_{nl}(k)$ in
eq.~(\ref{eq:Pdiskr}) by $r^{cl}_{n}\alpha_{nl}(k)$ and omitting
terms proportional to $N^{-1}$ and higher-order corrections, we rewrite
eq.~(\ref{eq:Pdiskr}) as
 \begin{equation}
   P \ = \ \left(2N+l+\frac{3}{2}\right)\,
   \frac{\alpha_{N+1,\,l} (k)-\alpha_{Nl} (k)}{\alpha_{N+1,\,l} (k)}\ .
  \label{eqPdiskr}
  \end{equation}
%Using the same considerations as above,
It is easy to show that  the
right-hand side of eq.~(\ref{eqPdiskr}) is just
%quasiclassical limit $N \to \infty$, the $P$-matrix $P={\cal P}_N$ where
the discrete analogue of the $P$-matrix $P$ defined as
\begin{equation}
  {\cal P}_N  \equiv  r^{cl}_{N}\
    \frac{1}{\alpha_{N+1,\,l} (k)}\;
       \frac{\Delta \alpha_{N+1,\,l} (k)}{\Delta r^{cl}_{N+1}}
         \label{disaP}\ .
\end{equation}
%\par
%We shall refer to the $P$-matrix defined by eq. (\ref{eq:Pdis3})
%at the distance $r$ equal to the natural channel radius $b_0$ as to the
%{\em discrete analogue of the $P$-matrix} ${\cal P}_{N}$.
%Really, using eq. (\ref{eq:cnin})
%we can rewrite (\ref{eq:Pdis3}) in the following form:
%In eq. (\ref{eq:Pdiskr}), the multiplier $(2N+l+5/2)$ plays the role of
%the channel radius in the oscillator representation, the function
%$a_{N+1,\,l}^{as}(k)$ represents the wave function in the oscillator
%representation on the truncation boundary, and the numerator is just
%the finite difference of  $a_{N+1,\,l}^{as}(k)$, i.e. eq.
%(\ref{eq:Pdiskr}) is just the finite difference analogue of eq.
%(\ref{eq:Popr}).
\par
Concluding this section, we have derived expressions that can be
used for calculation of $R$- or $P$-matrix in any variational approach
based on the harmonic oscillator expansions, e.g., in the shell model
approach. We have shown that if only the channel radius is equal to the
natural channel radius or any other radius $b_i$ fitting eq.
(\ref{frac}) or eq. (\ref{eq:ka}), the  variational energies coincide
with the $P$-matrix poles; in the quasiclassical limit,
% $N\to\infty$,
the $P$-matrices $P$ and ${\bar P}$ for the natural channel radius reduce
to their discrete analogues ${\cal P}_N$ and ${\bar {\cal P}}_N$,
respectively. These results link the inner structure of the
system with scattering characteristics and do not depend on the
interaction in the system, on the structure of the energy spectrum,
etc.
\section{Coulomb interaction within HORSE formalism.}
\subsection{Theory}\label{C-HORSE-Th}
\par
 We have discussed above only scattering on short-range potentials and
have  not allowed for the Coulomb interaction. Till now the only method of
treating the scattering of charged particles
within the HORSE approach is the one that has been proposed by
Kiev group in ref.~\cite{Okhr}. Matrix elements of the
Coulomb potential in the oscillator representation, $V^{Coul}_{nn'}$,
decrease very slowly as $n$ or/and $n'$ increases. Thus, following the
lines of ref. \cite{Okhr}, to allow for the Coulomb interaction within
the HORSE formalism one should increase the truncation boundary $N$
up to the values $N\sim 50$. In the case of long-range
Coulomb interaction, analytic expressions like  (\ref{eq:Snl}) and
(\ref{eq:Cnl}) for
the linearly independent solutions of the Schr\"{o}dinger equation in
the asymptotic region $n>N$ are unknown, and to calculate the solutions
one should use asymptotic expressions \cite{Okhr}. Using the method
of ref. \cite{Okhr}, it is possible to calculate with high accuracy
scattering observables in the case of few-channel scattering of charged
particles, but application of the method to the case of a large number of
channels is not feasible.
\par
The $P$-matrix discussed in the previous section can be used for formulation
of a version of the HORSE formalism allowing for the Coulomb
interaction that is free from the shortcomings of the method proposed in
ref. \cite{Okhr}.
\par
Let us consider the scattering in the system with the
interaction described by the potential $V=V^{Nucl}+V^{Coul}$, where
$V^{Nucl}$ is a short-range nuclear potential, and
$V^{Coul}=Z_{1}Z_{2}e^{2}/r$ is the Coulomb potential. In this case
instead of (\ref{Ylm}) the wave function is expressed as
\begin{equation}
 \Psi^{\pm}_{\bf k}({\bf r}) = \frac{4\pi}{k}\sum_{l,m}
  i^{l} e^{\pm i(\delta_l + \eta_l)}
  u_{l}(k,r) Y^{*}_{lm}(\Omega_{\bf k})  Y_{lm}(\Omega_{\bf r})\ ,
 \label{YlmC}
 \end{equation}
where $\eta_l = \arg \Gamma (1+l+i\zeta)$ is the Coulomb phase shift, and
Sommerfeld parameter $\zeta = Z_{1}Z_{2}e^{2}\mu /k$. We introduce the
channel radius $b$ that
is supposed to be large enough to neglect the nuclear potential $V^{Nucl}$
at the distances $r \geq b$, i.e., we suppose that $b \geq R_{Nucl}$
where $R_{Nucl}$ is
the range of $V^{Nucl}$. In the asymptotic region $r\geq b$, the partial
amplitudes $u_l (k,r)$ are expressed by eq.~(\ref{eq:assN1}) with
${\hat F}_{l}(k,r)=\frac{1}{r}\:\sqrt{\frac{\pi}{2v}} F_{l}(\zeta,kr)$ and
${\hat G}_{l}(k,r)= -\:\frac{1}{r}\: \sqrt{\frac{\pi}{2v}} G_{l}(\zeta,kr)$,
where regular $F_{l}(\zeta,kr)$ and irregular $G_{l}(\zeta,kr)$
Coulomb functions are defined according to ref.~\cite{GW}. The $P$-matrix,
$P^C$, accounting for the Coulomb asymptotics, satisfies the general
expression (\ref{genP}).
%wave function is of the form:
%  \begin{equation}
%   u_{l} (k,r)=\frac{1}{\sqrt{v}}\, e^{\delta_{l}^{Coul}}\left[
%    \cos \delta _{l}\, F_{l}(\zeta,kr)
%          \ +\ \sin \delta _{l}\, G_{l}(\zeta,kr)\right] \ ,
%  \label{eq:ascoul}
%  \end{equation}
%where $F_{l}(\zeta,kr)$ and $G_{l}(\zeta,kr)$ are regular and irregular
%Coulomb functions \cite{Abr}, respectively, $\delta_{l}^{Coul}$ is the
%Coulomb phase shift,  Sommerfeld parameter
%$\zeta = Z_{1}Z_{2}e^{2}m /k$, $Z_{1}e$ and $Z_{2}e$ are charges of the
%colliding particles. The Coulomb $P$-matrix, $P^C$, satisfies eq.
%(\ref{genP}) with ${J}_{l}(k,b) = F_{l}(\zeta,kr)$ and
%${N}_{l}(k,b) = G_{l}(\zeta,kr)$.
\par
Let us introduce an auxiliary short-range potential $V^{Sh}$ by cutting
Coulomb potential $V^{Coul}$ at the point $r=b$, i.e.
  \begin{equation}
  V^{Sh} = \left\{ \begin{array}{cl}
  V = V^{Nucl} + V^{Coul},  & \mbox{ $r \leq b$} \\
  0,                   & \mbox{ $r > b$}
  \end{array} \right. \,;  \\
  \ \   b \geq R_{Nucl}\ .
  \label{eq:pot1}
  \end{equation}
The continuum spectrum wave function corresponding to the potential $V^{Sh}$
is given by (\ref{Ylm}) with $\delta_{l} = \delta^{Sh}_{l}$, where
$\delta^{Sh}_{l}$ are the $l$-wave scattering phase shifts for the potential
$V^{Sh}$. In the asymptotic region $r\geq b$, the corresponding partial
amplitudes $u^{Sh}_{l}(k,r)$ are of the form:
   \begin{equation}
  u^{Sh}_{l}(k,r)\  =\ \cos \delta_{l}^{Sh}\, {f}_{l}(k,r) \ +\
           \sin \delta^{Sh}_{l}\,{g}_{l}(k,r) \ ,
  \label{eq:asm}
  \end{equation}
where ${f}_{l}(k,r)=\frac{k}{\sqrt{v}}\, {j}_{l}(kr)$ and ${g}_{l}(k,r)=
-\frac{k}{\sqrt{v}}\,{n}_{l}(kr)$.  The Schr\"{o}dinger
equation for the short-range potential $V^{Sh}$ can be
solved within the HORSE formalism, and the corresponding $P$-matrix,
$P^{Sh}$, can be calculated by eq. (\ref{Pgen}).
\par
An important point is that the $P$-matrix, i.e., the logarithmic derivative
of the wave function $u_{l}(k,r)$ at the point $r=b$, is dictated by the
interaction region $r\leq b$ only, and does not depend on the potential
in the asymptotic region $r >b$. Thus, $P$-matrices $P^C$ and $P^{Sh}$
corresponding to the potentials $V$ and $V^{Sh}$, respectively, should
be equal at the point $r=b$,
    \begin{equation}
    P^C = P^{Sh}.
   \label{eqCSh}
   \end{equation}
Substituting $P^C$ and $P^{Sh}$ in eq. (\ref{eqCSh}) by the right-hand
sides of eqs. (\ref{genP}) and (\ref{Pgen}), respectively, we
derive the following expression for the phase shift $\delta_l$
corresponding to the potential $V=V^{Nucl} + V^{Coul}$:
  \begin{equation}
  \tan \delta_l = \frac{ \left[ C_{Nl}(k)\,-\,
   {\cal G}_{NN}^{Sh}C_{N+1,\, l}(k)\right]W_{b}(j_{l},F_{l})\,+\,
    \left[ S_{Nl}(k)\,-\,
   {\cal G}_{NN}^{Sh}S_{N+1,\, l}(k)\right]W_{b}(n_{l},F_{l}) }
           { \left[ C_{Nl}(k)\,-\,
   {\cal G}_{NN}^{Sh}C_{N+1,\, l}(k)\right]W_{b}(j_{l},G_{l})\,+\,
    \left[ S_{Nl}(k)\,-\,
   {\cal G}_{NN}^{Sh}S_{N+1,\, l}(k)\right]W_{b}(n_{l},G_{l}) }\ .
  \label{tangen}
  \end{equation}
Here, ${\cal G}_{NN}^{Sh}$ is %the matrix element
defined by eq.~(\ref{eq:oscrm}) and corresponds to the Hamiltonian
with the auxiliary short-range potential~$V^{Sh}$,
the~quasi-Wronskian~$W_{b}(j_{l},F_{l})\equiv$
$\left. \left\{ \frac{d}{dr}[j_{l}(kr)]\,
F_{l}(\zeta,kr)\,-\,j_{l}(kr)\,\frac{d}{dr}F_{l}(\zeta,kr)\right\}
\right|_{r=b}$, and $W_{b}(n_{l},F_{l})$, $W_{b}(j_{l},G_{l})$ and
$W_{b}(n_{l},G_{l})$ are expressed similarly.
\par
Eq. (\ref{tangen}) defines within the HORSE formalism  the phase shift
$\delta_l$ in the case of  scattering of charged particles. It involves
diagonalization of the Hamiltonian with short-range interaction
$V^{Sh}$ only. Thus, moderate values of the truncation boundary $N$
can be used; as a result, the
corresponding algorithm appears to be much more effective as compared to
the method of ref. \cite{Okhr}.
%, and can be applied to multichannel calculations.
\par
Taking into account eq. (\ref{eq:oscph}), it is easy to obtain from
(\ref{tangen}) the expression for the phase shift $\delta_l$ in terms of the
phase shift $\delta_{l}^{Sh}$ corresponding to the auxiliary short-range
potential $V^{Sh}\:$:
  \begin{equation}
  \tan \delta_l = \frac{ W_{b}(j_{l},F_{l})\,-\,
   W_{b}(n_{l},F_{l})\tan \delta_{l}^{Sh} }
    { W_{b}(j_{l},G_{l})\,-\,W_{b}(n_{l},G_{l})\tan \delta_{l}^{Sh} }\ .
  \label{tantan}
  \end{equation}
This expression has been proposed in ref. \cite{Vin} for momentum-space
calculations of charged particle scattering.
\par
The partial amplitude $u_{l}^{Sh}(k,r)$ corresponding to the potential
$V^{Sh}$ obtained by the HORSE approach, perfectly matches in the
interaction region $r\leq b$ the required
wave function $u_{l}(k,r)$ corresponding to the potential
$V=V^{Nucl}+V^{Coul}$, but $u_{l}^{Sh}(k,r)$ and $u_{l}(k,r)$ differ
%essentially
at larger distances where the asymptotics~(\ref{eq:assN1}) with
${\hat F}_{l}(k,r)=\frac{1}{r}\:\sqrt{\frac{\pi}{2v}} F_{l}(\zeta,kr)$ and
${\hat G}_{l}(k,r)= -\:\frac{1}{r}\: \sqrt{\frac{\pi}{2v}} G_{l}(\zeta,kr)$
should be used instead of (\ref{eq:asm}). However, typically only the
interaction region is of importance for calculations
of electromagnetic transition probabilities and other
observables. At the same time, it
should be taken into account, that asymptotics are only significant in
calculation of normalization factors for continuum spectrum wave functions.
Thus, before calculating matrix elements of operators of physical
observables, $u_{l}^{Sh}(k,r)$ should be renormalized. The
renormalization factor ${\cal N}$ can be easily derived from the  equation
$u_{l}(k,b)={\cal N}u_{l}^{Sh}(k,b)$ where $u_{l}(k,r)$
and $u_{l}(k,r)$ are given by (\ref{eq:asm}) and
(\ref{eq:assN1}), respectively:
  \begin{equation}
  {\cal N}\,=\,\frac
  {\cos \delta_{l}\,F_{l}(\zeta,kb)\,-\,
     \sin \delta _{l}\,G_{l}(\zeta,kb)}
  {\cos \delta_{l}^{Sh}\,j_{l}(kb)\,-\,
      \sin \delta_{l}^{Sh}\, n_{l}(kb)}\ .
  \label{eq:N}
\end{equation}
\subsection{Numerical illustration.}
\par
We illustrate the applicability if our method for treating Coulomb
interaction within the HORSE formalism by calculations of the
single-channel proton-nucleus potential scattering. We use
Woods-Saxon potential with conventional
surface spin-orbit term as the nuclear potential, $V^{Nucl}$.
Parameters of $V^{Nucl}$ and of the Coulomb potential $V^{Coul}$
taken from ref.~\cite{Coupl2} correspond to the $p$--$^{15}$N
scattering. Most of the result presented below were obtained with the
oscillator level spacing parameter $\hbar\omega=18$~MeV.
We made use of Lanczos smoothing of potential energy matrix
elements \cite{Hung}
that improves the convergence. Note that the convergence of
the results of HORSE calculations
%as $N$ increases
without Coulomb interaction is usually achieved at $N \approx 5\div 6$.
%; most of the
%results presented below are obtained with the truncation boundary $N=9$
%(i.e., the truncated Hamiltonian matrix $H^{l}_{nn'}$ used in
%calculations is $10\times 10$ matrix).
\par
Channel radius $b$ used for construction of the auxiliary
potential $V^{Sh}$, is a free parameter of the method. It should be
taken larger than the range of the Woods-Saxon potential $R_{Nucl}$. On
the other hand, the truncated Hamiltonian matrix $\tilde{H}_{nn'}^l$
should carry information about the jump of potential $V^{Sh}$ at the
point $r=b$. Thus, $b$ should be chosen less than the classical turning
point, $r_{N}^{cl}=2r_{0}\,\sqrt{N+l/2+3/4}$, of the
oscillator function $R_{Nl}(r)$, i.e.\ of the function with the
largest range in the set of oscillator
functions $\left\{R_{nl}(r),\ n\leq N\right\}$ used for the
construction of the truncated Hamiltonian matrix $\tilde{H}_{nn'}^l$.
Figure~3 gives the $s$-wave phase shifts
$\delta_0$ corresponding to the proton energies $E=2$ and 10~MeV
calculated with different values of $b$.
%Straight horizontal lines on  the fig.~3
%represent the exact values of $\delta_0$.
In our case, $R_{Nucl}\approx 5.5$~fm while $\hbar\omega=18$~MeV
corresponds $r_0\approx 1.5$~fm and hence $N=9$ corresponds
$r_{N}^{cl}\approx 9$~fm.
It is seen from fig.~3 that
the plot of $\delta_{0}(b)$ has a plateau between $R_{Nucl}$ and
$r_{N}^{cl}$ that reproduces well the exact values of $\delta_{0}$.
Moderate variations of $\hbar\omega$  cause only very slight changes of
the phase shift while the increase of $N$ results only in
extension of the plateau to larger values of $b$. The phase shifts
$\delta_{l}^{j}(b)$ with $l>0$ also have the plateaus that reproduce
the exact values even better than in the case $l=0$.
\par
So, one should assign to the channel radius $b$ a
value from the interval $R_{Nucl}<b<r_{N}^{cl}$. Results presented below
were obtained with $b=7$~fm.
\par
Plots of the $s$-wave phase shift $\delta_{0}$ versus proton
energy $E$ are presented in fig.~\ref{fig4}. The exact values of
$\delta_{0}$ were obtained by numerical integration of the
Schr\"{o}dinger equation in the differential form.
It is seen that the results obtained in our approach exactly reproduce
the phase shift up to $E=30$~MeV and only the phase shift obtained
with $N$ as small as 4 deviate a little from the exact ones. We note
that there is a sharp resonance in the $s$-wave scattering. The
resonance is generated by the Coulomb barrier and disappears if the
Coulomb interaction is turned off. The phase shift in the vicinity of
the resonance depends crucially on the resonant energy. Our calculations
reproduce well the position of the resonance. It is
interesting that the best description of the resonant energy is obtained
in calculations with $N=4$, and hence the $N=4$ phase shift is closer
to the exact one in the vicinity of the resonance than the phase shifts
obtained with larger values of $N$. We note, however, that the
height of the jump of the phase shift in the vicinity of the resonance
is reproduced in calculations with larger values of $N$ better than in
calculations with $N=4$.
\par
For comparison, the phase shifts calculated by the method suggested by
Kiev group~\cite{Okhr}, are also plotted in fig.~\ref{fig4}. The
thin solid curve with diamonds was obtained with the truncation
boundaries recommended in ref.~\cite{Okhr}: the Hamiltonian matrix was
truncated at $N=70$ and hence $71\times 71$ matrix was diagonalized in
this single-channel problem, however the matrix of the short-range
nuclear potential was truncated at $N^{Sh}=50$. It is seen that the Kiev
method describes the phase shifts with the same accuracy that ours
(note, however, small deviation of the Kiev phase shift from the exact
one in the low-energy region) but is much less efficient in applications
because it involves diagonalization of a large matrix. If the truncation
boundaries of the Kiev method are reduced essentially, the
description of the phase shift becomes worse. This is illustrated in
fig.~\ref{fig4} by dots which present the phase shift obtained with
$N^{Sh}=8$ and $N=70$, i.e.\ the short-range potential was truncated to
$9\times 9$ matrix while the Coulomb interaction was truncated
to $71\times 71$ matrix, and therefore in calculation of the phase
shift it was needed to diagonalize the $71\times 71$ Hamiltonian matrix.
Note that contrary to the Kiev method, we obtain an excellent description
of the phase shift with $N=8$, i.e.\ diagonalizing the $9\times 9$
Hamiltonian matrix.  If the matrix of the same rank is diagonalized in
the Kiev method, i.e.\ the truncation boundary $N$ is reduced by an
order of magnitude, the method fails in the description of the Coulomb
asymptotics of the wave functions, of the phase shifts, etc.
\par
The scattering phase shift is a very important but not the only
characteristic of the continuum spectrum states. In calculation of
electromagnetic transition probabilities and other observables, one
needs matrix elements of various operators. Within the HORSE formalism,
the matrix elements are expressed through the standard matrix elements
in the oscillator basis (\ref{eq:oscrad})  of the
operator of interest and
the wave function in oscillator representation $a_{nl}(k)$ entering
expansion (\ref{eq:row}). The plots of $a^2_{nl}(k)$ for $l=0$ and
$n=0,\,1,\,2$ are given
in fig.~\ref{fig5}. The exact values of $a^2_{nl}(k)$ were calculated as
        \begin{equation}
   a_{nl}^{2}(q)=\left[ \int_{0}^{\infty}u_{l}(k,r)\,R_{nl}(r)\;r^2dr
   \right]^2
   \label{int}
   \end{equation}
where the wave function $u_{l}(k,r)$ was obtained
by numerical integration of the Schr\"{o}dinger equation in the
differential form. It is seen that the wave function in oscillator
representation $a_{nl}(k)$ is reproduced with the same high accuracy in
our method as the phase shift. The description of $a_{nl}(k)$ in
the Kiev method is also approximately of the same accuracy as the one of
the phase shift. The most important though small enough deviation from
the exact result is seen only in the vicinity of the resonance and is
dictated by the resonant energy that differs a little in various
approximations. Note that the shape of the resonant curves is well
reproduced for all values of truncation boundary. The best description
of the resonant energy is obtained accidentally in the $N=4$
calculation, and the $N=4$ curves are closer to the exact results in the
vicinity of the resonance than the ones obtained with larger values of
$N$. Note, however, that the integral $\int a^2_{nl}(k)\;dE$ in the
vicinity of the resonance is better reproduced in $N>4$ calculations.
\par
The coordinate space radial wave functions $\tilde u_l(kr)$ are
reconstructed within the HORSE formalism as
\begin{equation}
\tilde u_{l}(k,r) = r\sum_{n=0}^{M} a_{nl}(k)\: R_{nl}(r)\ .
\label{reconstruct}
\end{equation}
The plots of $\tilde u_l(kr)$ are given in fig.~\ref{fig6}. It is seen
that the wave function deviates from the exact one if $M=N$
even in the case when the truncation boundary $N$ is large enough to
reproduce exactly the phase shifts. The deviation is easily eliminated
by setting $M$ to be large enough and allowing in the expansion
(\ref{reconstruct}) for the wave function in oscillator representation
in the asymptotic region $a_{nl}^{as}(k)$ with $n>N$ that is calculated
by a simple formula (\ref{eq:asosc}). It is interesting to note that in
the great majority of nuclear applications like calculations of
electromagnetic transition rates and of other observables, allowing for
$a_{nl}^{as}(k)$ with $n>N$ does not effect the results, and hence the
deviation of the wave function $\tilde u_l(kr)$ obtained by
(\ref{reconstruct}) with $M=N$ from the exact one that seems to be
essential, is really of no importance.
\par
The $p$-wave and $d$-wave phase shifts and corresponding $a^2_{nl}(k)$
are plotted on figs.~\ref{fig7} and \ref{fig8}.  It is seen that the
$p$-wave and $d$-wave observables are reproduced in our approach with
the same accuracy as the $s$-wave ones.
\par
Concluding this section, we developed the Coulomb HORSE approach that
is efficient in calculations of continuum spectrum wave functions and
scattering observables within the HORSE method in the case of short
range~+~Coulomb interaction.

\section{Multichannel scattering.}
\label{Mult-Ch-Sect}
\par
In this section we generalize the results of previous sections to the
multichannel case.
\par
In the multichannel case, the wave function
is of the form \cite{GW,Baz,Taylor}:
\begin{equation}
      \Psi_{\tilde\Gamma_{i}}^{(\pm)}\ =
     \frac{4\pi}{k_{\Gamma_{i}}}
   \sum_{\mathop{\nu, J, s_{\Gamma}, }
   \limits_{l_{\Gamma}, l_{\Gamma _{i}}}}\,
i^{l_{\Gamma}}e^{\pm i\eta_{l_\Gamma}}\,
   \langle \Omega _{r_{\Gamma}} l_{\Gamma} s_{\Gamma} \sigma_{\Gamma}
   \left| {\cal Y}^{J} \right|
   \Omega _{k_{\Gamma_{i}}} l_{\Gamma_{i}} s_{\Gamma_{i}} \sigma_{\Gamma_{i}}
    \rangle\;
   \Phi_{\nu  l_{\Gamma} s_{\Gamma}}(\chi_{\Gamma})\;
   u_{\Gamma (\Gamma _{i})}^{J(\pm)}\left(k_{\Gamma},r_{\Gamma}\right)\ .
    \label{mwf}
\end{equation}
Expression (\ref{mwf}) can be used both for charged or neutral
particles, in the latter case the Coulomb phase shifts
$\eta_{l_\Gamma}=0$.
It is supposed that the incident plane (or Coulomb-distorted) wave
in the case of $\Psi_{\tilde\Gamma_{i}}^{(+)}$ or the outgoing
plane (or Coulomb-distorted) wave  in the case of
$\Psi_{\tilde\Gamma_{i}}^{(-)}$ is present only in the ``physical''
channel labeled by the multiindex $\tilde\Gamma_i=\{\nu,
\mbox{\boldmath $k$}_{\Gamma_i}, s_{\Gamma_i},\sigma_{\Gamma_i}\}$,
while outgoing waves in the case of $\Psi_{\tilde\Gamma_{i}}^{(+)}$ and
ingoing waves in the case of $\Psi_{\tilde\Gamma_{i}}^{(-)}$ are present
in all ``physical'' channels $\tilde\Gamma=\{\nu,
\mbox{\boldmath $k$}_{\Gamma}, s_{\Gamma},\sigma_{\Gamma}\}$.
Here, $s_{\Gamma}$ and $\sigma_{\Gamma}$ are the total spin of the
colliding particles in the channel $\tilde\Gamma$ and its projection,
respectively; $\mbox{\boldmath $k$}_{\Gamma}$ is the relative motion
momentum in the channel $\tilde\Gamma$; all the rest quantum numbers in
the channel $\tilde\Gamma$, in particular the ones related to the
internal state of the colliding particles described by the internal wave
function $\Phi_{\nu  l_{\Gamma} s_{\Gamma}}(\chi_{\Gamma})$, are
labeled by the index $\nu$. In the right-hand side of eq.~(\ref{mwf})
we use multiindex
$\Gamma = \{\nu, J, l_{\Gamma}, s_{\Gamma},\sigma_{\Gamma} \}$
to label channels characterized by a definite value of the total
angular momentum $J$, $l_{\Gamma}$ is the orbital angular momentum in
the channel $\Gamma$. The transformation of the states of the
channels $\Gamma$ with a definite $J$  into the states of the
``physical" channels $\tilde\Gamma$ characterized by the vector of the
relative motion momentum $\mbox{\boldmath $k$}_{\Gamma}$ of the incident
or outgoing plane (or Coulomb-distorted) wave, is performed with the
help of (see, e.g., \cite{GW})
\begin{eqnarray}
   \langle \Omega _{r_{\Gamma}} l_{\Gamma} s_{\Gamma} \sigma_{\Gamma}
   \left| {\cal Y}^{J} \right|
   \Omega _{k_{\Gamma_{i}}} l_{\Gamma_{i}} s_{\Gamma_{i}} \sigma_{\Gamma_{i}}
    \rangle & = &
   \sum_{m_{\Gamma}, m_{\Gamma_{i}}, M}
    Y_{l_{\Gamma}m_{\Gamma}}\left(\Omega_{r_{\Gamma}}\right)
    Y^{*}_{l_{\Gamma_{i}}m_{\Gamma_{i}}}\left(\Omega_{k_{\Gamma_{i}}}\right)
           \nonumber  \\
  & \times &
   \langle l_{\Gamma} m_{\Gamma} s_{\Gamma} \sigma_{\Gamma}
   \left| JM \right. \rangle
   \langle l_{\Gamma_{i}} m_{\Gamma_{i}} s_{\Gamma_{i}} \sigma_{\Gamma_{i}}
   \left| JM \right. \rangle \ ,
                    \label{angpart}
\end{eqnarray}
where $\langle lm s\sigma\left| JM\right.\rangle$ stands for a
Clebsch-Gordan coefficient, $m_{\Gamma}$ is the projection of the
orbital angular momentum $l_\Gamma$ in the channel $\Gamma$, and $M$ is
the projection of the total angular momentum $J$.
\par
The relative motion in the channel $\Gamma$ is described by the radial
wave function
$u^{J(\pm)}_{\Gamma (\Gamma_{i})}\left( k_{\Gamma},r_{\Gamma}\right)$.
In the case when the incident plane wave corresponds to the channel
$\tilde\Gamma_{i}$ and outgoing spherical
waves are present in all channels, the radial channel functions
%normalized to the unit flux
are of the form in the asymptotic region $r_{\Gamma} \ge b_{\Gamma}\,$:
\begin{equation}
    u_{\Gamma (\Gamma_{i})}^{J(+)}\left( k_{\Gamma},r_{\Gamma}\right) \
    = \frac{1}{2i} \left(
        \hat{H}^{(+)}_{l_{\Gamma}}\left( k_{\Gamma},r_{\Gamma}\right)\
            S_{\Gamma \Gamma_{i}}\ -\
      \hat{H}^{(-)}_{l_{\Gamma}}\left( k_{\Gamma},r_{\Gamma}\right)
\ \delta_{\Gamma \Gamma_{i}} \right),
                \label{muas}
\end{equation}
where $S_{\Gamma \Gamma^{\prime}}$ is a matrix element of the $S$-matrix,
\begin{equation}
 \hat{H}^{(\pm)}_{l}( k,r)\ \equiv  \hat{G}_{l}( k,r)\ \pm
  i\,\hat{F}_{l}( k,r),
%%\ \mathop{\longrightarrow}\limits_{r\to \infty} \
  %%\frac{1}{r\sqrt{v}} \exp{[\pm i(kr-\pi l/2)]} \ ,
                \label{Has}
\end{equation}
and $\hat{F}_{l}( k,r)$ and $\hat{G}_{l}( k,r)$ are defined in sect.~III.
In calculations of photodisintegration cross sections one needs
radial channel functions
$u_{\Gamma (\Gamma_{o})}^{J(-)}\left( k_{\Gamma},r_{\Gamma}\right)=
\left(u_{\Gamma (\Gamma_{o})}^{J(+)}\left(
k_{\Gamma},r_{\Gamma}\right)\right)^*$
characterized by the outgoing plane (or Coulomb-distorted) wave in
the channel $\tilde\Gamma_o$ and ingoing spherical waves in all
channels.
%
%In the asymptotic region $r_{\Gamma} \ge b_{\Gamma}\,$,
%\begin{equation}
%    u_{\Gamma (\Gamma_{o})}^{J(-)}\left( k_{\Gamma},r_{\Gamma}\right) \
%    = \frac{1}{2i} \left(
%        \hat{H}^{(+)}_{l_{\Gamma}}\left( k_{\Gamma},r_{\Gamma}\right)\
%\ \delta_{\Gamma \Gamma_{o}}\ -\
%      \hat{H}^{(-)}_{l_{\Gamma}}\left( k_{\Gamma},r_{\Gamma}\right)
%            S_{\Gamma \Gamma_{o}} \right)
%                \label{mumas}
%\end{equation}
%
Below we shall discuss the radial channel functions
$u^{J(+)}_{\Gamma (\Gamma_{i})}\left( k_{\Gamma},r_{\Gamma}\right)$
only; of course, the same formalism can be applied to the functions
$u^{J(-)}_{\Gamma (\Gamma_{o})}\left( k_{\Gamma},r_{\Gamma}\right)$.
\par
It is convenient to introduce radial channel function matrix
(we suppose that the number of open channels is $M$ and use square
brackets to denote $M\times M$ matrices in the channel space),
\begin{equation}
    \left[ u^{J(+)}\left( k, r \right) \right] \ \equiv \
    \left(
   \begin{array}{cccc}
      u_{1 (1)}^{J(+)}\left( k_{1},r_{1}\right)
    &  u_{1 (2)}^{J(+)}\left( k_{1},r_{1}\right) & \cdots
    &  u_{1 (M)}^{J(+)}\left( k_{1},r_{1}\right)  \\
     u_{2 (1)}^{J(+)}\left( k_{2},r_{2}\right)
    & u_{2 (2)}^{J(+)}\left( k_{2},r_{2}\right)  & \cdots
    & u_{2 (M)}^{J(+)}\left( k_{2},r_{2}\right)  \\
   \multicolumn{4}{c}{\makebox[7cm]{\dotfill} } \\
     u_{M (1)}^{J(+)}\left( k_{M},r_{M}\right)
    &  u_{M (2)}^{J(+)}\left( k_{M},r_{M}\right) & \cdots
    &  u_{M (M)}^{J(+)}\left( k_{M},r_{M}\right)
        \end{array}   \right)\ .
           \label{umatr}
\end{equation}
A natural generalization of eq.~(\ref{eq:Popr}) leads to the following
definition of a multichannel $P$-matrix  \cite{Baz,Jaffe}:
% \newcounter{abc}
 \addtocounter{equation}{1}
\setcounter{abc}{\value{equation}}
\setcounter{equation}{0}
\renewcommand{\theequation}{\arabic{abc}\alph{equation}}
\begin{eqnarray}
%\begin{equation}
%     \left[ P \right] \ = \ \left[ b \right]
%      \left[ \left. \frac{du^{(+)}(k,r)}{dr}\right|_{r=b} \right]
%     \left[ u^{(+)}( k, b ) \right]^{-1} .
%                \label{Pmatr}
%\end{equation}
     \left[ P \right] \ &  & = \ \left[ b \right]
%   \left[ \left. \frac{du^{(+)}(k,r)}{dr}\right|_{r=b} \right]
     \left[ u^{J(+)}(k,b) \right]^{\prime}
     \left[ u^{J(+)}( k, b ) \right]^{-1}
                \label{Pmatr}           \\
   & & = \ \left. \left. \left[ b \right]
 \left\{
    \left[ \hat{H}^{(-)} \left( k, b \right) \right]^{\prime} -
\left[\hat{H}^{(+)}\left( k,b \right)\right]^{\prime}\left[ S\right]
    \right\}
 %  \left\{
  \right\{
 \left[ \hat{H}^{(-)} \left( k, b \right) \right] -
  \left[\hat{H}^{(+)}\left(k,b\right)\right]\left[S\right]\right\}^{-1}.
            \label{du}
\end{eqnarray}
The symmetry of the multichannel $P$-matrix $\left[ P \right]$ is
discussed in the Appendix.
In eqs.~\mbox{(\ref{Pmatr})--(\ref{du})},
\renewcommand{\theequation}{\arabic{equation}}
\setcounter{equation}{\value{abc}}
\begin{eqnarray}
\lefteqn{
    \left[ u^{J(+)}\left( k, b \right) \right]^{\prime} \equiv }&
              \nonumber\\
& \phantom{\displaystyle \int u dr}
\renewcommand{\arraystretch}{1.7}
   \left(
   \begin{array}{cccc}
\displaystyle\left.
\frac{u_{1(1)}^{J(+)}\left(k_{1},r_{1}\right)}{dr_1}\right|_{r_1=b_1}
&\displaystyle\left.
\frac{u_{1(2)}^{J(+)}\left(k_{1},r_{1}\right)}{dr_1}\right|_{r_1=b_1}
           & \cdots
    &\displaystyle\left.
\frac{u_{1(M)}^{J(+)}\left(k_{1},r_{1}\right)}{dr_1}\right|_{r_1=b_1}
                                       \\
\displaystyle\left.
\frac{u_{2 (1)}^{J(+)}\left( k_{2},r_{2}\right)}{dr_2}\right|_{r_2=b_2}
    & \displaystyle\left.
\frac{u_{2 (2)}^{J(+)}\left( k_{2},r_{2}\right)}{dr_2}\right|_{r_2=b_2}
             & \cdots
    & \displaystyle\left.
\frac{u_{2 (M)}^{J(+)}\left( k_{2},r_{2}\right)}{dr_2}\right|_{r_2=b_2}
                                            \\
   \multicolumn{4}{c}{\makebox[7cm]{\dotfill} } \\
     \displaystyle\left.
\frac{u_{M(1)}^{J(+)}\left( k_{M},r_{M}\right)}{dr_M}\right|_{r_M=b_M}
    &     \displaystyle\left.
\frac{u_{M(2)}^{J(+)}\left( k_{M},r_{M}\right)}{dr_M}\right|_{r_M=b_M}
                 & \cdots
    &     \displaystyle\left.
\frac{u_{M(M)}^{J(+)}\left( k_{M},r_{M}\right)}{dr_M}\right|_{r_M=b_M}
        \end{array}   \right)\ ,
           \label{umatrprime}
\end{eqnarray}
$\left[ S \right]$ is the
$S$-matrix, the diagonal channel radius matrix
\begin{equation}
    \left[ b \right] \ \equiv \
    \left(
   \begin{array}{cccc}
     b_1  &     &   & 0 \\
          & b_2 &   &    \\
          &     & \ddots & \\
      0   &     &   & b_M
        \end{array}   \right)\, ,
           \label{bmatr}
\end{equation}
%and for  the matrices
%$\left[ \left. \frac{du^{(+)}(k,r)}{dr}\right|_{r=b} \right]$ and
%$\left[ u^{(+)}( k, b ) \right]$ one can use the following expressions:
%\begin{equation}
%   \left[ u^{(+)}\left( k, b \right) \right] \ = \
%    \left[ H^{(-)} \left( k, b \right) \right] \ -\
%        \left[ H^{(+)} \left( k,b \right) \right] \left[ S \right]
%                \label{u-h}\, ,
%\end{equation}
%and
%\begin{equation}
%    \left[ \left. \frac{du^{(+)}(k,r)}{dr}\right|_{r=b} \right] \ = \
%    \left[ H^{(-)} \left( k, b \right) \right]^{\prime} \ -\
%        \left[ H^{(+)} \left( k,b \right) \right]^{\prime} \left[ S \right]
%                \label{du-dh}\, ,
%\end{equation}
%where $\left[ S \right]$ is $S$-matrix,
and diagonal matrices
\begin{equation}
    \left[ \hat{H}^{(\pm)}\left( k,b \right) \right] \ \equiv \
    \left(
   \begin{array}{cccc}
     \hat{H}_{l_{1}}^{(\pm)}\left( k_{1},b_{1} \right) &     &   & 0 \\
  & \hat{H}_{l_{2}}^{(\pm)}\left( k_{2},b_{2} \right) &   &    \\
          &     & \ddots & \\
 0 & & & \hat{H}_{l_{M}}^{(\pm)}\left( k_{M},b_{M} \right)
        \end{array}   \right)   %\, ,
           \label{Hmatr}
\end{equation}
and
\begin{equation}
    \left[ \hat{H}^{(\pm)} \left( k, b \right) \right]^{\prime}
   \equiv    \left(
         \renewcommand{\arraystretch}{1.3}
 \begin{array}{cccc}
\displaystyle
    \left. \frac{d\hat{H}_{l_1}^{(\pm)} \left( k_{1},\,r_{1} \right)}
       {dr_1} \right|_{r_{1}=b_1}
                                     &     &   & 0 \\
   & \displaystyle \left.
 \frac{d\hat{H}_{l_2}^{(\pm)}\left( k_{2},\,r_{2}
                     \right)}{dr_2}\right|_{r_{2}=b_2}
                                     &   &    \\
          &     & \ddots &  \\
      0   &     &   & \displaystyle \left.
  \frac{d\hat{H}_{l_M}^{(\pm)}\left( k_{M},\,r_{M}
                       \right)}{dr_M}\right|_{r_{M}=b_M}
        \end{array}   \right)\, .
           \label{dHmatr}
\end{equation}
\par
Within the multichannel HORSE formalism \cite{YaFi,SmSh},
the radial channel functions
$u^{J(+)}_{\Gamma (\Gamma_{i})}\left( k_{\Gamma},r_{\Gamma}\right)$ are
expanded in terms of the harmonic oscillator functions,
  \begin{equation}
    u^{J(+)}_{\Gamma (\Gamma_{i})}\left( k_{\Gamma},r_{\Gamma}\right)\ =
\ \sum_{n=0}^{\infty} a_{n \Gamma (\Gamma_{i})}\left( k_{\Gamma}\right)
          R_{nl_{\Gamma}}\left( r_{\Gamma}\right) \
  \label{mexp}
  \end{equation}
[we omit the indexes $J$ and $(\pm)$ in notations of the wave functions
in oscillator representation:
$a_{n \Gamma (\Gamma_{i})}\left(k_{\Gamma}\right)\equiv
a^{J(\pm)}_{n \Gamma (\Gamma_{i})}\left( k_{\Gamma}\right)\;$].
It is convenient to expand also the internal channel wave functions
$\Phi_{\nu  l_{\Gamma} s_{\Gamma}}(\chi_{\Gamma})$ in oscillator
function series, but generally any other representation for the
functions $\Phi_{\nu  l_{\Gamma} s_{\Gamma}}(\chi_{\Gamma})$ can be
used. The matrix of the projectile--target interaction within the
multichannel HORSE formalism is truncated according to
eq.~(\ref{trunc}) in each particular channel $\Gamma$ independently from
the others, and as a result we shall get a set of truncation boundaries
$\left\{N_{\Gamma}\right\}$.
%; different values can be used for the truncation
%boundaries $N_{\Gamma}$ in different channels $\Gamma$.
In the asymptotic region $n \ge N_{\Gamma}$, the wave function in
oscillator representation is of the form
\begin{equation}
    a^{as}_{n \Gamma (\Gamma_{i})}\left( k_{\Gamma}\right) \ = \
     \delta_{\Gamma \Gamma_{i}}
      C^{(-)}_{nl_{\Gamma}}\left( k_{\Gamma} \right)\ -\
     S_{\Gamma \Gamma_{i}} C^{(+)}_{nl_{\Gamma}}\left( k_{\Gamma} \right)\ ,
  \label{anG}
\end{equation}
where $C^{(\pm)}_{nl}( k) = C_{nl}( k) \pm iS_{nl}( k)$.
\par
Truncating the
Hamiltonian matrix $H^{\Gamma \Gamma^{\prime}}_{nn^{\prime}}$ at
$n=N_{\Gamma}$ in each  particular channel $\Gamma$
and diagonalizing the truncated Hamiltonian
matrix $\tilde{H}^{\Gamma \Gamma^{\prime}}_{nn^{\prime}}$,
one can calculate matrix elements
  \begin{equation}
  {\cal G}^{\Gamma \Gamma^{\prime}}_{nn'} = -\sum_{\lambda}\,
  \frac{ \left( \gamma_{\lambda n}^{\Gamma}\right)^{*}\,
       \gamma_{\lambda n'}^{\Gamma^{\prime}} }
  { E_{\lambda}\,-\,E }\,T_{n',\,n'+1}^{\Gamma^{\prime}}\ ,
  \label{muoscrm}
  \end{equation}
which we use to construct the matrix
\begin{equation}
    \left[ {\cal G} \right] \ \equiv \
    \left(
   \begin{array}{cccc}
  {\cal G}^{1\,1}_{N_{1}\,N_{1}} & {\cal G}^{1\,2}_{N_{1}\,N_{2}}
       & \cdots & {\cal G}^{1\,M}_{N_{1}\,N_{M}} \\
  {\cal G}^{2\,1}_{N_{2}\,N_{1}} & {\cal G}^{2\,2}_{N_{2}\,N_{2}}
       & \cdots & {\cal G}^{2\,M}_{N_{2}\,N_{M}} \\
   \multicolumn{4}{c}{\makebox[4cm]{\dotfill} } \\
  {\cal G}^{M\,1}_{N_{M}\,N_{1}} & {\cal G}^{M\,2}_{N_{M}\,N_{2}}
       & \cdots & {\cal G}^{M\,M}_{N_{M}\,N_{M}}
        \end{array}   \right)\ .
           \label{muGmatr}
\end{equation}
In eq.~(\ref{muoscrm}) $T_{n,\,n+1}^{\Gamma}$ is the relative motion kinetic
energy matrix element for the channel $\Gamma$, and
$E_{\lambda}$ and $\gamma_{\lambda n}^{\Gamma}$ are eigenvalues
and corresponding eigenvectors of the matrix
$\tilde{H}^{\Gamma \Gamma^{\prime}}_{nn^{\prime}}$.
Equation~(\ref{muoscrm}) is a generalization of eq.~(\ref{eq:oscrm})
to the multichannel case. The matrix elements
${\cal G}^{\Gamma \Gamma^{\prime}}_{nn'} $ are useful for calculations
of the multichannel wave function in the harmonic oscillator
representation in the interaction region $n \le N_{\Gamma}$ in any
particular channel $\Gamma$ by
  \begin{equation}
  a_{n \Gamma (\Gamma_{i})}\left( k_{\Gamma}\right)\  = \
   \sum_{\Gamma^{\prime}}
   {\cal G}^{\Gamma \Gamma^{\prime}}_{nN_{\Gamma^{\prime}}}\;
   a^{as}_{N_{\Gamma^{\prime}}+1,\, \Gamma^{\prime}\, (\Gamma_{i})}
      \left( k_{\Gamma^{\prime}}\right)\ ,
  \label{mucnin}
  \end{equation}
while eq.~(\ref{anG}) should be used for calculation of
$a^{as}_{n \Gamma (\Gamma_{i})}\left( k_{\Gamma}\right)$ in the asymptotic
regions $n\ge N_{\Gamma}$. To make use of
eq.~(\ref{anG}), one should first calculate the $S$-matrix. Within the
multichannel HORSE formalism it can be done by the following expression
\cite{SmSh}:
  \begin{equation}
  \left[ S \right]\ =\ \left\{ \left[ C_{N}^{(+)}(k)\right]\ -\
  \left[ {\cal G} \right]\ \left[ C_{N+1}^{(+)}(k)\right] \right\}^{-1}
  \ \left\{ \left[ C_{N}^{(-)}(k)\right]\ -\
  \left[ {\cal G} \right]\ \left[ C_{N+1}^{(-)}(k)\right] \right\}\ ,
   \label{Smatr}
\end{equation}
or, taking into account the symmetry property of the $S$-matrix\footnote{The
$S$-matrix is defined with cuts in the complex energy plane starting from
threshold energies $\epsilon_\Gamma$ of all channels $\Gamma$. If the
threshold energies $\epsilon_\Gamma$ are different for different channels,
the symmetry property
%%%%  $\left[ S \right]=\left[ S \right]^{Tr}$
(\protect\ref{S-matr-sym}) may not hold
in the complex energy plane. In this case some of the formulas below should be
modified. However, all the formulas below can be used at energies $E$ such
that ${\rm Re}\;E>\max\;\{\epsilon_\Gamma\}$ or
${\rm Re}\;E<\min\;\{\epsilon_\Gamma\}$. \label{S-sym-foot} },
%%%%%%%%%%%
  \begin{equation}
       \left[ S \right]=\left[ S \right]^{Tr}
 \label{S-matr-sym}
       \end{equation}
where the index $Tr$ labels a transposed matrix,
  \begin{equation}
  \left[ S \right]\ =\ \left\{ \left[ C_{N}^{(-)}(k)\right]\ -\
  \left[ C_{N+1}^{(-)}(k)\right]\ \left[ {\cal G} \right]^{Tr} \right\}
  \ \left\{ \left[ C_{N}^{(+)}(k)\right]\ -\
  \left[ C_{N+1}^{(+)}(k)\right]\ \left[ {\cal G} \right]^{Tr} \right\}^{-1} \ .
   \label{TrSmatr}
\end{equation}
Here diagonal matrices
$\left[ C_{N+1}^{(\pm)}(k)\right]$ and $\left[ C_{N}^{(\pm)}(k)\right]$
are defined as
\begin{equation}
\left[ C_{N+1}^{(\pm)}(k)\right] \ \equiv \
    \left(
   \begin{array}{cccc}
  C^{(\pm)}_{N_{1}+1,\,l_{1}}\left( k_{1} \right) &  &  & 0 \\
  & C^{(\pm)}_{N_{2}+1,\,l_{2}}\left( k_{2} \right)   &  &  \\
          &     & \ddots & \\
      0   &     &   &  C^{(\pm)}_{N_{M}+1,\,l_{M}}\left( k_{M} \right)
        \end{array}   \right)
           \label{CN+1}
\end{equation}
and
\begin{equation}
\left[ C_{N}^{(\pm)}(k)\right] \ \equiv \
    \left(
   \begin{array}{cccc}
  C^{(\pm)}_{N_{1}\,l_{1}}\left( k_{1} \right) &  &  & 0 \\
  & C^{(\pm)}_{N_{2}\,l_{2}}\left( k_{2} \right)   &  &  \\
          &     & \ddots & \\
      0   &     &   &  C^{(\pm)}_{N_{M}\,l_{M}}\left( k_{M} \right)
        \end{array}   \right)\ .
           \label{CN}
\end{equation}
\par
In order to define the $P$-matrix within the multichannel HORSE
formalism, we substitute the $S$-matrix in eq.~(\ref{du}) by the
right-hand side of eq.~(\ref{TrSmatr}). After some algebra, we derive:
  \begin{equation}
  \begin{array}{rl}
 \left[P\right] \  = \  &\left[b\right] \left\{
  \left[\hat{H}^{(+)}(k,b)\right]^{\prime}\left[C^{(-)}_{N}(k)\right]-
  \left[\hat{H}^{(-)}(k,b)\right]^{\prime}\left[C^{(+)}_{N}(k)\right]
    \right.   \\
    & -  \left.
  \left(\left[\hat{H}^{(+)}(k,b)\right]^{\prime}\left[C^{(-)}_{N+1}(k)\right] -
  \left[\hat{H}^{(-)}(k,b)\right]^{\prime}\left[C^{(+)}_{N+1}(k)\right]\right)
          \left[{\cal G}\right]^{Tr} \right\}   \\
     & \times  \left\{
  \left[\hat{H}^{(+)}(k,b)\right]\left[C^{(-)}_{N}(k)\right] -
  \left[\hat{H}^{(-)}(k,b)\right]\left[C^{(+)}_{N}(k)\right]
    \right.  \\
    & -   \left.
  \left(\left[\hat{H}^{(+)}(k,b)\right]\left[C^{(-)}_{N+1}(k)\right] -
  \left[\hat{H}^{(-)}(k,b)\right]\left[C^{(+)}_{N+1}(k)\right]\right)
          \left[{\cal G}\right]^{Tr} \right\}^{-1} .
  \end{array}
   \label{P1}
  \end{equation}
$P$-matrix is known \cite{Baz} to
be a real matrix. Nevertheless, most of the matrices entering (\ref{P1})
are complex ones. It is useful for applications to rewrite eq.~(\ref{P1})
in a form that does not contains complex variables:
  \begin{equation}
  \begin{array}{rl}
 \left[P\right] \  = \  &\left[b\right] \left\{
  \left[\hat{F}(k,b)\right]^{\prime}\left[C_{N}(k)\right]-
  \left[\hat{G}(k,b)\right]^{\prime}\left[S_{N}(k)\right]
    \right.   \\
    & -  \left.
  \left(\left[\hat{F}(k,b)\right]^{\prime}\left[C_{N+1}(k)\right] -
  \left[\hat{G}(k,b)\right]^{\prime}\left[S_{N+1}(k)\right]\right)
          \left[{\cal G}\right]^{Tr} \right\}   \\
     & \times  \left\{
  \left[\hat{F}(k,b)\right]\left[C_{N}(k)\right] -
  \left[\hat{G}(k,b)\right]\left[S_{N}(k)\right]
    \right.  \\
    & -   \left.
  \left(\left[\hat{F}(k,b)\right]\left[C_{N+1}(k)\right] -
  \left[\hat{G}(k,b)\right]\left[S_{N+1}(k)\right]\right)
          \left[{\cal G}\right]^{Tr} \right\}^{-1} ,
  \end{array}
   \label{P2}
  \end{equation}
where the matrices $ \left[\hat{F}(k,b)\right]$, $ \left[\hat{G}(k,b)\right]$,
$\left[\hat{F}(k,b)\right]^{\prime}$, $\left[\hat{G}(k,b)\right]^{\prime}$,
$\left[S_{N}(k)\right]$, $\left[C_{N}(k)\right]$, $\left[S_{N+1}(k)\right]$,
and $ \left[C_{N+1}(k)\right]$  are diagonal matrices with matrix elements
equal to $\hat{F}_{l_{\Gamma}}( k_{\Gamma},b_{\Gamma})$,
$\hat{G}_{l_{\Gamma}}( k_{\Gamma},b_{\Gamma})$,
$\displaystyle\left. \frac{d}{dr_\Gamma}\hat{F}_{l_{\Gamma}}
( k_{\Gamma},r_{\Gamma})\right|_{r_{\Gamma}=b_{\Gamma}}$,
$\displaystyle \left. \frac{d}{dr_\Gamma}
\hat{G}_{l_{\Gamma}}(k_{\Gamma},r_{\Gamma})
\right|_{r_{\Gamma}=b_{\Gamma}}$,
$S_{N_{\Gamma}\,l_{\Gamma}}( k_{\Gamma} )$,
$C_{N_{\Gamma}\,l_{\Gamma}}( k_{\Gamma} )$,
$S_{N_{\Gamma}+1,\,l_{\Gamma}}( k_{\Gamma} )$, and
$C_{N_{\Gamma}+1,\,l_{\Gamma}}( k_{\Gamma} )$, respectively, e.g.,
\begin{equation}
     \left[\hat{F}(k,b)\right] \ = \ \left(
   \begin{array}{cccc}
  \hat{F}_{l_{1}}( k_{1},b_{1})  & & & 0 \\
  & \hat{F}_{l_{2}}( k_{2},b_{2})  &  &  \\
          &     & \ddots & \\
      0   &     &   &  \hat{F}_{l_{M}}( k_{M},b_{M})
        \end{array}   \right) \ .
           \label{fl}
\end{equation}

\par
Eq.~(\ref{P2}) generalizes eq.~(\ref{Pgen}) to the multichannel case. It
can be used for the calculation of the multichannel $P$- or $R$-matrix,
$\left[R\right]=\left[P\right]^{-1}$, within the harmonic oscillator
expansion method for any set of the channel radii
$\left\{b_{\Gamma}\right\}$ large enough
to neglect the potential energy at the distances $r_{\Gamma}\ge b_{\Gamma}$
in each channel $\Gamma$.
\par
$P$-matrix poles coincide with eigenenergies of the truncated Hamiltonian
matrix $\tilde{H}^{\Gamma \Gamma^{\prime}}_{nn^{\prime}}$ if the matrix
$\left[{\cal G}\right]^{Tr}$ is not actually present in the inverse
matrix in the right-hand side of eqs.(\ref{P1})--(\ref{P2}), i.e. when
the condition
\begin{equation}
   \left[\hat{F}(k,b)\right]\left[C_{N+1}(k)\right]\ = \
  \left[\hat{G}(k,b)\right]\left[S_{N+1}(k)\right]
    \label{bcondit}
\end{equation}
is satisfied. Equation~(\ref{bcondit}) is a matrix equation that should
be solved to find a set of channel radii $\left\{b_{\Gamma}\right\}$.
However, all the
matrices entering eq.~(\ref{bcondit}) are actually diagonal matrices.
Thus, eq.~(\ref{bcondit}) reduces to a set of $M$ uncoupled
equations (\ref{frac}) each corresponding to some channel $\Gamma$, and the
channel radius $b_{\Gamma}$ for any particular channel $\Gamma$ can
be found independently from the others. As a result, the set of
{\em natural channel radii} $\left\{b^{0}_{\Gamma}\right\}$ can be
introduced with the help of eq.~(\ref{eq:a}) applied to all channels
$\Gamma$ in turn. Of course, the natural channel radii $b^{0}_{\Gamma}$
are defined with the same accuracy as in the
single-channel case.
\par
If all of the channel radii $b_{\Gamma}$ are set to be equal to the
corresponding natural channel radii $b^{0}_{\Gamma}$ [or, more generally,
if the set of the channel radii $\left\{b_{\Gamma}\right\}$ is any solution of
eq.~(\ref{bcondit})], the general expression  for the $P$-matrix
(\ref{P2}) reduces to
  \begin{equation}
\renewcommand{\arraystretch}{1.3}
\begin{array}{rl}
 \left[P\right]   =  & \displaystyle \frac{2}{\hbar}
 \left[b^{0}\right] \left[T\right]
    \left\{ \left[\hat{G}(k,b^{0})\right]^{\prime} \left[S_{N}(k)\right] -
       \left[\hat{F}(k,b^{0})\right]^{\prime}\left[C_{N}(k)\right] \right.   \\
   & - \left. \left( \left[\hat{G}(k,b^{0})\right]^{\prime}
       \left[S_{N+1}(k)\right] -
        \left[\hat{F}(k,b^{0})\right]^{\prime}\left[C_{N+1}(k)\right] \right)
     \left[{\cal G}\right] \right\}
           \left[\hat{G}(k,b^0)\right]^{-1}  \left[C_{N+1}(k)\right] \ ,
          \end{array}
   \label{Pb1}
  \end{equation}
where $\left[ T \right]$ is a diagonal matrix built up by
kinetic energy matrix elements
$T_{N_{\Gamma},\, N_{\Gamma}+1}^{\Gamma}$. Eq.~(\ref{Pb1}) is a multichannel
generalization of eq.~(\ref{Pi}). In the same approximation that was
used to derive eq.~(\ref{eq:Posc1}), i.e., in the case when  one
can use approximations
$\displaystyle \hat{F}_{l_{\Gamma}}( k_{\Gamma},r_{\Gamma})\approx
\frac{1}{r_\Gamma\sqrt{v_\Gamma}}
\sin \left({k_\Gamma r_\Gamma-\frac{\pi l_\Gamma}{2}}\right)$
and $\displaystyle \hat{G}_{l_{\Gamma}}( k_{\Gamma},r_{\Gamma})\approx
\frac{1}{r_\Gamma\sqrt{v_\Gamma}}
\cos \left({k_\Gamma r_\Gamma-\frac{\pi l_\Gamma}{2}}\right)$
in every channel $\Gamma$ at the distances $r_{\Gamma}\ge b_{\Gamma}$,
eq.~(\ref{Pb1}) can be simplified to
  \begin{equation}
\begin{array}{r@{}l}
 \left[P\right] \ = \  &-\left[b^{0}\right] \left[k\right]
\left[T\right] \Bigl\{ \left[S_{N}(k)\right] \left[S_{N+1}(k)\right]
    +   \left[C_{N}(k)\right] \left[C_{N+1}(k)\right]   \\
   & -  \left( \left[S_{N+1}(k)\right]^2 +
           \left[C_{N+1}(k)\right]^2 \right)
    \left[\hat{G}(k,b^{0})\right] \left[C_{N+1}(k)\right]^{-1}
    \left[{\cal G}\right]
    \left[\hat{G}(k,b^{0})\right]^{-1}  \left[C_{N+1}(k)\right]  \Bigr\}
         - \left[ \hat{1} \right] ,
           \end{array}
   \label{Pb2}
  \end{equation}
where $\left[ k \right]$ is the diagonal  matrix of the channel momenta
$k_{\Gamma}$, and $\left[ \hat{1} \right]$ is the unit matrix.

\par
Further simplification of eq.~(\ref{Pb2}) can be performed making
use of asymptotic expressions (\ref{eq:Snlass}) and (\ref{eq:Cnlass}) of
the functions $S_{nl}(k)$ and $C_{nl}(k)$. As a result we get:
  \begin{equation}
 \left[P\right] \ = \  2\left[\sqrt{(N+1)\left(N+l+\frac{3}{2}\right)}\ \right]
     \left(  \left[\beta\right] -
  \left[b^{0} \right]^{-\frac{1}{2}}  \left[r_{0} \right]^{-1}
  \left[{\cal G}\right]
  \left[r_{0} \right]^{+1}  \left[b^{0} \right]^{\frac{1}{2}}
         \right)
    \ -\ \left[ \hat{1} \right] \ .
   \label{Pb3}
  \end{equation}
Here $\left[ r_0\right]$,
$\left[\vphantom{\int\limits_a^a}\sqrt{(N+1)(N+l+\frac{3}{2})}\right]$
and $\left[\beta\right]$ are the diagonal matrices with non-zero matrix
elements equal to
$r^{\Gamma}_{0}=\sqrt{\hbar / (\mu_\Gamma \omega )}$ where $\mu_\Gamma$ is
the reduced mass in the channel $\Gamma$,
$\sqrt{(N_{\Gamma}+1)(N_{\Gamma}+l_{\Gamma}+\frac{3}{2})}$ and
 \begin{equation}
 \beta_{\Gamma} =
 \left( \frac{2N_{\Gamma} +l_{\Gamma}+7/2}
 {2N_{\Gamma}+l_{\Gamma}+3/2} \right)^{1/4}
 \cos \left\{ 2k_{\Gamma}r_{0} \left(
 \sqrt{N_{\Gamma}+\frac{l_{\Gamma}}{2}+\frac{7}{4}} -
 \sqrt{N_{\Gamma}+\frac{l_{\Gamma}}{2}+\frac{3}{4}}\right)\right\}  ,
 \label{mbeta}
 \end{equation}
respectively. A multichannel
generalization of eq.~(\ref{eq:Pdis3}) can be derived from  eq.~(\ref{Pb3})
in the limit $N_{\Gamma} \to \infty$:
  \begin{equation}
 \left[P\right] \ = \  2\left[N+\frac{l}{2}+\frac{5}{4} \right]
     \left( \left[ \hat{1} \right]  -
  \left[b^{0} \right]^{-\frac{1}{2}} \left[r_{0} \right]^{-1}
  \left[{\cal G}\right]
  \left[r_{0} \right]  \left[b^{0} \right]^{\frac{1}{2}}
         \right)
    \ -\ \left[ \hat{1} \right]  .
   \label{Pb4}
  \end{equation}
where $\left[N+\frac{l}{2}+\frac{5}{4} \right]$
is a diagonal matrix with
$\left(N_{\Gamma}+\frac{l_{\Gamma}}{2}+\frac{5}{4} \right)$ standing for
the non-zero matrix elements.
\par
Equation~(\ref{Pb4}) defines the {\em multichannel discrete
analogue of the $P$-matrix}. Note that if the reduced mass $\mu_\Gamma$
is the same in all channels $\Gamma$, the matrix $\left[r_{0}\right]$
appears to be proportional to the unit matrix. In this case expression
(\ref{Pb4}) simplifies to
  \begin{equation}
 \left[P\right] \ = \  2\left[N+\frac{l}{2}+\frac{5}{4} \right]
     \left( \left[ \hat{1} \right]  -
  \left[b^{0} \right]^{-\frac{1}{2}}
  \left[{\cal G}\right]
    \left[b^{0} \right]^{\frac{1}{2}}
         \right)
    \ -\ \left[ \hat{1} \right]  .
   \label{Pb4-mueq}
  \end{equation}
A further simplification of eq.~(\ref{Pb4-mueq}) is possible
if the channel radius $b^{0}_{\Gamma}$ is the same in all
channels $\Gamma$ (for example, if truncation boundaries $N_\Gamma$ and
orbital angular momenta $l_\Gamma$ are the same in all channels
$\Gamma$), because in this case $\left[b^{0}\right]^{-\frac{1}{2}}
\left[{\cal G}\right]\left[b^{0}\right]^{\frac{1}{2}} =
\left[{\cal G}\right]$.
\par
As in the single-channel case, being derived in the quasiclassical limit
$N\to\infty$, the multichannel discrete analogue of the $P$-matrix
provides a very accurate approximation to the exact $P$-matrix even for
very small values of truncation boundaries $N_\Gamma$. This is not
surprising because only the matrix $\left[{\cal G}\right]$ entering
eqs.~(\ref{Pb4}) and (\ref{Pb4-mueq}) mixes the channels, all the rest
matrixes in eqs.~(\ref{Pb4}) and (\ref{Pb4-mueq}) are diagonal in the
channel space and provide in each channel the same accuracy as in the
single-channel case.
   \par
The above multichannel $P$-matrix formalism can be used to derive
a multichannel generalization of our Coulomb--HORSE method.

\par
Consider a multichannel scattering in the system with the interaction
%in channel $\Gamma$
described  by a  superposition of a short-range nuclear potential
$V^{Nucl}_{\Gamma\Gamma^{\prime}}$ and Coulomb potential
$V^{Coul}_{\Gamma\Gamma}$.  Contrary to the Coulomb interaction, the
nuclear potential $V^{Nucl}_{\Gamma\Gamma^{\prime}}$ couples the channels.
The wave function is given by eq.~(\ref{mwf}) where
$\eta_{l_{\Gamma}}= \arg \Gamma(1+l_{\Gamma}+i\zeta_{\Gamma})$
and $\zeta_{\Gamma}$ are the Coulomb phase shift and Sommerfeld
parameter, respectively,
in the channel $\Gamma$.
In the asymptotic region $r_\Gamma\ge b_\Gamma$, the radial channel
functions
$u^{J(+)}_{\Gamma (\Gamma_{i})}\left( k_{\Gamma},r_{\Gamma}\right)$
are of the form:
\begin{equation}
    u^{J(+)}_{\Gamma (\Gamma_{i})}\left( k_{\Gamma},r_{\Gamma}\right) \
    = \frac{1}{2i} \left(
        \hat{G}^{(+)}_{l_{\Gamma}}\left( k_{\Gamma},r_{\Gamma}\right)\
            S_{\Gamma \Gamma_{i}}\ -\
      \hat{G}^{(-)}_{l_{\Gamma}}\left( k_{\Gamma},r_{\Gamma}\right)
\ \delta_{\Gamma \Gamma_{i}} \right)  ,
                \label{Cmuas}
\end{equation}
where
\begin{eqnarray}
 \hat{G}^{(\pm)}_{l}( k,r) & \equiv &
  -\frac{1}{r} \sqrt{ \frac{\pi}{2v} } G_{l}(\zeta, kr)\ \pm
  i\,\frac{1}{r} \sqrt{ \frac{\pi}{2v} } F_{l}(\zeta, kr)
               \label{CHas}           \\ \nonumber
  & \mathop{\longrightarrow}\limits_{r\to \infty} \ &
  \frac{1}{r\sqrt{v}}
  \exp{\left[\pm i\left(kr - \zeta \ln{2kr} - \frac{\pi l}{2} + \eta_{l}
         \right) \right]}\vphantom{^{\displaystyle\int}} \:  ,
                \label{CHas-ass}
\end{eqnarray}
and $F_{l}(\zeta , kr)$ and $G_{l}(\zeta , kr)$ are the regular and irregular
Coulomb function, respectively.
%The partial amplitudes eq.~(\ref{Cmuas})
%form the matrix of radial channel functions
%$\left[ u^{(+)}\left( k, r \right) \right]$ of the type as eq.(\ref{umatr}).
\par
By analogy with section \ref{C-HORSE-Th},
%the case of  potential scattering,
we introduce an auxiliary
short-range potential $V^{Sh}_{\Gamma}$ by cutting Coulomb
potential $V^{Coul}_{\Gamma}$ at distances $r_\Gamma=b_{\Gamma}$ in the each
channel $\Gamma$ (it is supposed that $b_\Gamma\ge R^{Nucl}_\Gamma$
where $R^{Nucl}_\Gamma$ is the range of the nuclear interaction in the
corresponding channel):
  \begin{equation}
  V^{Sh}_{\Gamma\Gamma^{\prime}} = \left\{ \begin{array}{cl}
  V_{\Gamma\Gamma^{\prime}} = V^{Nucl}_{\Gamma\Gamma^{\prime}} +
\delta_{\Gamma\Gamma^{\prime} }V^{Coul}_{\Gamma\Gamma},  &
  \mbox{ $r_\Gamma \leq b_{\Gamma}$} \\
  0,                   & \mbox{ $r_\Gamma > b_{\Gamma}$}
  \end{array} \right. \,;  \\
\quad  \ \   b_{\Gamma} \geq R_\Gamma^{Nucl}\ .
  \label{mpot1}
  \end{equation}
%We shall denote the matrix of radial wave function for this potential
%as $\left[ u^{(+)Sh}\left( k, r \right) \right]$. Because the short-range
%character of modified potential,
In the asymptotic region $r_\Gamma\ge b_\Gamma$, the radial channel
functions
$u^{J(+){Sh}}_{\Gamma (\Gamma_{i})}\left( k_{\Gamma},r_{\Gamma}\right)$
obtained with the  auxiliary potential (\ref{mpot1}) are given by
eq.~(\ref{muas}) with an auxiliary $S$-matrix
$S^{Sh}_{\Gamma\Gamma_i}$ standing for $S_{\Gamma\Gamma_i}$:
\begin{equation}
    u_{\Gamma (\Gamma_{i})}^{J(+)Sh}\left( k_{\Gamma},r_{\Gamma}\right) \
    = \frac{1}{2i} \left(
        \hat{H}^{(+)}_{l_{\Gamma}}\left( k_{\Gamma},r_{\Gamma}\right)\
            S_{\Gamma \Gamma_{i}}^{Sh}\ -\
      \hat{H}^{(-)}_{l_{\Gamma}}\left( k_{\Gamma},r_{\Gamma}\right)
\ \delta_{\Gamma \Gamma_{i}} \right) \ .
                \label{Shmuas}
\end{equation}
\par
As in the case of single-channel scattering, for the set of channel
radii $\{b_\Gamma\}$ we have an equation
   \begin{equation}
    \left[ P^C \right]\ =\ \left[ P^{Sh} \right],
   \label{meqCSh}
   \end{equation}
where $ \left[ P^C \right]$ is the $P$-matrix of the
multichannel Coulomb problem and  the $P$-matrix $\left[ P^{Sh} \right]$
corresponds to the   auxiliary multichannel problem with short-range
interaction (\ref{mpot1}).
The coupled channel equations for this  auxiliary multichannel problem
can be solved within the  multichannel HORSE formalism, and the
corresponding scattering matrix $S^{Sh}$, the matrix of channel functions
$\left[ u^{J(+)Sh}\left( k, r \right) \right]$ and the $P$-matrix $P^{Sh}$,
can be calculated.
\par
The Coulomb $P$-matrix  $ \left[ P^C \right]$ can be substituted in
eq.~(\ref{meqCSh}) by its expression through
the multichannel Coulomb $S$-matrix $\left[ S \right]$ given by
eq.~(\ref{du}) with $\left[ \hat{H}^{(\pm )} \left( k, b \right) \right]$
replaced by
diagonal matrices  $\left[ \hat{G}^{(\pm )} \left( k, b \right) \right]$
defined according to the general rule (\ref{fl}). As a result, we can
derive the following expression for the $S$-matrix of the multichannel
Coulomb problem:
  \begin{equation}
  \begin{array}{ccl}
  \left[ S \right] & = &
  \left\{
  \left[ b \right]^{-1}\, \left[ P^{Sh} \right]\,
  \left[ \hat{G}^{(+)}(k,b)\right]\ -\
  \left[ \hat{G}^{(+)}(k,b)\right]^{\prime}
  \right\}^{-1} \\ \nonumber
  & \times & \left\{
  \left[ b \right]^{-1}\, \left[ P^{Sh} \right]\,
  \left[ \hat{G}^{(-)}(k,b)\right]\ -\
  \left[ \hat{G}^{(-)}(k,b)\right]^{\prime}
  \right\}
\end{array}
   \label{ShSmP}
\end{equation}
[cf. with the general expression of the multichannel $S$-matrix through the
$P$-matrix ~(\ref{S-through-P}) given in the Appendix].
Equation~(\ref{P2}) can be used to calculate  $\left[ P^{Sh} \right]$ entering
(\ref{ShSmP}).
\par
It is also possible to derive an expression of the multichannel Coulomb
$S$-matrix $\left[ S \right]$ in terms of the auxiliary $S$-matrix
$\left[ S^{Sh} \right]$. Using the symmetry of the $P$-matrix [see
eq.~(\ref{P-sym2}) in the Appendix] we rewrite (\ref{meqCSh}) as
\begin{equation}
\left[\frac{\mu}{b}\right]^{-1} \left[P^C\right]
  =  \left[P^{Sh}\right]^{Tr}
\left[\frac{\mu}{b}\right]^{-1}  .
\label{PC-PSh-sym}
\end{equation}
We substitute  $\left[P^C\right]$ and $\left[P^{Sh}\right]$
in (\ref{PC-PSh-sym}) by their expressions (\ref{du}) through
$\left[ S \right]$ and $\left[ S^{Sh} \right]$,
respectively, to obtain
\begin{equation}
\begin{array}{@{}lr}
\lefteqn{\displaystyle
   \left[\frac{\mu}{b}\right]^{-1}   \left[b\right]
   \left\{ \left[\hat{G}^{(-)}(k,b)\right]^{\prime}-
  \left[\hat{G}^{(+)}(k,b)\right]^{\prime}  \left[S\right]\right\}
   \left\{ \vphantom{ \left[\hat{G}^{(-)}(k,b)\right]^{\prime}}
      \left[\hat{G}^{(-)}(k,b)\right]
  -\left[\hat{G}^{(+)}(k,b)\right] \left[S\right]\right\}^{-1} }\\[2mm]
 &\displaystyle   =
   \left\{  \vphantom{ \left[\hat{G}^{(-)}(k,b)\right]^{\prime}}
        \left[\hat{H}^{(-)}(k,b)\right]-
   \left[S^{Sh}\right] \left[\hat{H}^{(+)}(k,b)\right] \right\}^{-1}
   \left\{
  \left[\hat{H}^{(-)}(k,b)\right]^{\prime}-
   \left[S^{Sh}\right] \left[\hat{H}^{(+)}(k,b)\right]^{\prime}
  \right\} \left[b\right] \left[\frac{\mu}{b}\right]^{-1}   .
  \end{array}
\label{SthrougSSh-au}
\end{equation}
Solving equation (\ref{SthrougSSh-au}) with respect to $\left[S\right]$ we
derive:
\begin{equation}
\begin{array}{rl}
  \left[S\right]  & = \displaystyle
   \left[\frac{\mu}{b}\right]^{-1} \left[b\right]
  \left(
 \left\{\left[\hat{H}^{(-)}(k,b)\right]\left[\hat{G}^{(-)}(k,b)\right]^{\prime}
   - \left[\hat{H}^{(-)}(k,b)\right]^{\prime}\left[\hat{G}^{(-)}(k,b)\right]
      \right\}  \right. \\[2mm]
  & \left. -  \left\{
    \left[\hat{H}^{(+)}(k,b)\right]\left[\hat{G}^{(-)}(k,b)\right]^{\prime}-
   \left[\hat{H}^{(+)}(k,b)\right]^{\prime}\left[\hat{G}^{(-)}(k,b)\right]
       \right\}  \left[S^{Sh}\right]
            \right) \\[2mm]
  & \times
  \left(  \left\{
   \left[\hat{H}^{(-)}(k,b)\right]\left[\hat{G}^{(+)}(k,b)\right]^{\prime}-
   \left[\hat{H}^{(-)}(k,b)\right]^{\prime}\left[\hat{G}^{(+)}(k,b)\right]
    \right\}  \right. \\[2mm]
 & \left.   - \left\{
   \left[\hat{H}^{(+)}(k,b)\right]\left[\hat{G}^{(+)}(k,b)\right]^{\prime}-
   \left[\hat{H}^{(+)}(k,b)\right]^{\prime}\left[\hat{G}^{(+)}(k,b)\right]
      \right\}  \left[S^{Sh}\right]
  \right)^{-1}
   \left[b\right]^{-1}\displaystyle \left[\frac{\mu}{b}\right] .
  \end{array}
\label{SthrougSSh}
\end{equation}
In the derivation we made use of
the symmetry of the $S$-matrix (\ref{S-matr-sym}) (note the footnote
\ref{S-sym-foot} on page~\pageref{S-sym-foot}).
\par
Equation~(\ref{SthrougSSh}) is a
multichannel generalization of eq.~(\ref{tantan}).
\par
As in the single-channel case [cf.\ eq.~(\ref{eq:N})], to obtain
the radial channel functions
$u^{J(+)}_{\Gamma (\Gamma_{i})}\left( k_{\Gamma},r_{\Gamma}\right)$
of the multichannel Coulomb problem in the interaction region
$r_\Gamma\le b_\Gamma$, one should renormalize the radial channel
functions
$u^{J(+){Sh}}_{\Gamma (\Gamma_{i})}\left( k_{\Gamma},r_{\Gamma}\right)$
obtained with the  auxiliary potential (\ref{mpot1}). The matrix of
radial channel functions $\left[ u^{J(+)}\left( k,r\right)\right]$  in
the interaction region can be calculated as
\begin{equation}
\left[ u^{J(+)}\left( k,r\right)\right] =
\left[ {\cal N}\right] \left[u^{J(+){Sh}}\left( k,r\right)\right],
\label{Mult-C-ren}
\end{equation}
where $\left[u^{J(+){Sh}}\left( k,r\right)\right]$ is the radial channel
function matrix of the auxiliary multichannel problem, and
the renormalization matrix $\left[ {\cal N}\right]$ is given by
\begin{equation}
  \left[ \cal{N} \right] =
  \left\{
  \left[ \hat{G}^{(+)} \left( k,b \right) \right] \left[ S \right]
  - \left[ \hat{G}^{(-)} \left( k, b \right) \right]
  \right\} \left\{
  \left[ \hat{H}^{(+)} \left( k,b \right) \right] \left[ S^{Sh} \right]
  - \left[ \hat{H}^{(-)} \left( k, b \right) \right]
                     \right\}^{-1} .
\label{Mult-C-ren-Matr}
\end{equation}
Note that the renormalization matrix $\left[ {\cal N}\right]$ is not
diagonal. This is because the long-range Coulomb
interaction modifies in a different manner radial channel functions in
different channels. As a
result, the relative weights of channel amplitudes in the interaction
region are changed by the Coulomb interaction. This requires some
rearrangement of the auxiliary wave function in the interaction region that
can be achieved only by mixing its components
$u^{J(+){Sh}}_{\Gamma (\Gamma_{i})}\left( k_{\Gamma},r_{\Gamma}\right)$,
which is generated by non-diagonal matrix elements of
$\left[ {\cal N}\right]$.

\par
In the asymptotic region $r_\Gamma\ge b_\Gamma$, the radial channel
functions
$u^{J(+)}_{\Gamma (\Gamma_{i})}\left( k_{\Gamma},r_{\Gamma}\right)$
are given by expression (\ref{Cmuas}).

\par
We suppose to present the results of applications of the
multichannel Coulomb--HORSE formalism to nuclear reactions in future
publications.

\section{Summary}

In this paper we have developed $P$- and $R$-matrix formalism within
harmonic oscillator representation of scattering theory. With the
help of this formalism, one can use well-developed $P$- or $R$-matrix
technique~\cite{Lane,Jaffe} in combination with standard variational
approaches based
on harmonic oscillator expansion~\cite{HOMPh}. In particular, one can
use the results of nuclear structure calculations obtained by standard
shell model codes to find resonance positions and widths, cross
sections of various nuclear reactions and other scattering
observables.
\par
We derived expressions for calculation
of $R$- or $P$-matrix at any distance from the origin covered by the set of
oscillator basis functions employed in the variational procedure. We have
shown that there is a natural channel radius associated with the given
oscillator basis. For the natural channel radius, the expressions for
the $P$- or $R$-matrix are simplified and a discrete analogue of the
$P$-matrix can be introduced. The discrete analogue of the
$P$-matrix is expressed through the finite difference of the wave
function in the oscillator representation in the same manner as the
usual $P$-matrix is expressed through the derivative of the wave
function in the coordinate representation. It has  the same properties
as the conventional $P$-matrix, in particular,  its poles are just the
eigenenergies obtained in variational approach with oscillator basis.
The discrete analogue was shown to
be equal to the $P$-matrix in the quasiclassical limit of large number
of basis functions, however, it gives a very good approximation for
the $P$-matrix  even in the case when only one oscillator basis function
is involved in the variational procedure.
\par
The discrete analogue of the
$P$-matrix is useful for calculation of scattering and
reactions of uncharged particles. It cannot be used for scattering
calculations in the case when Coulomb or another long-range
interaction is present in the open channel. However,  in this case one
can calculate the $P$- or $R$-matrix at shorter distances
than the natural channel radius for investigation of scattering
observables within the  variational oscillator-basis approach. For
such investigations, we developed a very efficient Coulomb--HORSE
formalism.
\par
In Section~\ref{Mult-Ch-Sect}, we generalized all our results on the
case of arbitrary number of channel.

\bigskip

%%%{\bf Acknowledgements}.
\acknowledgments
This work was supported in part by
Competitive Center at St.~Petersburg State University, State Program
``Russian Universities'' and
Russian Foundation of Basic Research. One of the authors (A.M.S.) would
like to thank for financial support and hospitality Niels Bohr
Institute where a part of this work was done as well as International
Institute of Theoretical and Applied Physics at
Iowa State University where this work was finished.

\appendix\section*{Symmetry of multichannel $P$-matrix.}
\renewcommand{\theequation}{A.\arabic{equation}}

The expression of the multichannel $P$-matrix through the $S$-matrix is
given by eq.~(\ref{du}) which can be used to express the $S$-matrix
through the $P$-matrix:
\begin{equation}
\begin{array}{rl}
 \left[S\right] \ & =
  \  \left\{ \left[b\right]^{-1}
  \left[P\right]\left[\hat{H}^{(+)}(k,b)\right]-
  \left[\hat{H}^{(+)}(k,b)\right]^{\prime} \right\}^{-1} \\
    & \times
  \  \left\{ \left[b\right]^{-1}
  \left[P\right]\left[\hat{H}^{(-)}(k,b)\right]-
  \left[\hat{H}^{(-)}(k,b)\right]^{\prime} \right\} .
  \end{array}
\label{S-through-P}
\end{equation}
Using the symmetry of the $S$-matrix (\ref{S-matr-sym}) (note the footnote
\ref{S-sym-foot} on page~\pageref{S-sym-foot}), we can rewrite
eq.~(\ref{S-through-P}) as
\begin{equation}
\begin{array}{rl}
 \left[S\right] \ & =
  \  \left\{ \left[\hat{H}^{(-)}(k,b)\right]
  \left[P\right]^{Tr}\left[b\right]^{-1} -
  \left[\hat{H}^{(-)}(k,b)\right]^{\prime} \right\} \\
    & \times
  \  \left\{ \left[\hat{H}^{(+)}(k,b)\right]
  \left[P\right]^{Tr}\left[b\right]^{-1} -
  \left[\hat{H}^{(+)}(k,b)\right]^{\prime} \right\}^{-1} \ .
  \end{array}
\label{Str-through-P}
\end{equation}
It follows from eqs.~(\ref{S-through-P})--(\ref{Str-through-P}) that
\begin{equation}
\begin{array}{@{}lr}
\lefteqn{
   \left\{ \left[b\right]^{-1}
  \left[P\right]\left[\hat{H}^{(+)}(k,b)\right]-
  \left[\hat{H}^{(+)}(k,b)\right]^{\prime} \right\}^{-1}
    \left\{ \left[b\right]^{-1}
  \left[P\right]\left[\hat{H}^{(-)}(k,b)\right]-
  \left[\hat{H}^{(-)}(k,b)\right]^{\prime} \right\}    }\\[2mm]
&  = \left\{ \left[\hat{H}^{(-)}(k,b)\right]
  \left[P\right]^{Tr}\left[b\right]^{-1} -
  \left[\hat{H}^{(-)}(k,b)\right]^{\prime} \right\}
   \left\{ \left[\hat{H}^{(+)}(k,b)\right]
  \left[P\right]^{Tr}\left[b\right]^{-1} -
  \left[\hat{H}^{(+)}(k,b)\right]^{\prime} \right\}^{-1}  ,
  \end{array}
\label{Semi-fabr}
\end{equation}
and after some algebra we obtain:
\begin{equation}
\begin{array}{r@{}l}
   \left\{
   \left[\hat{H}^{(+)}(k,b)\right]^{\prime}
\right.&\left.
                \left[\hat{H}^{(-)}(k,b)\right]
  - \left[\hat{H}^{(-)}(k,b)\right]^{\prime} \left[\hat{H}^{(+)}(k,b)\right]
   \right\}
  \left[P\right]^{Tr}\left[b\right]^{-1}    \\[2mm]
&  =
   \left[b\right]^{-1} \left[P\right]
   \left\{
   \left[\hat{H}^{(+)}(k,b)\right]^{\prime} \left[\hat{H}^{(-)}(k,b)\right]
  - \left[\hat{H}^{(-)}(k,b)\right]^{\prime} \left[\hat{H}^{(+)}(k,b)\right]
   \right\}  .
  \end{array}
\label{Wronsk-use}
\end{equation}

The Wronskian of the solutions $\hat{H}^{(+)}_{l}( k,r)$ and
$\hat{H}^{(-)}_{l}( k,r)$ defined by eq.~(\ref{Has}), is
\begin{equation}
{\hat{H}^{(+)^{\prime}}_{l}}( k,r)\:\hat{H}^{(-)}_{l}( k,r)
      - {\hat{H}^{(-)^{\prime}}_{l}}( k,r)\:\hat{H}^{(+)}_{l}( k,r)=
%%%%%  \frac{2ik}{r^2v} =
      \frac{2i}{\hbar}\ \frac{\mu}{r^2} \ .
\label{Wronskpm}
\end{equation}
Therefore eq.~(\ref{Wronsk-use}) appears to be equivalent to
\begin{equation}
\left[\frac{\mu}{b}\right] \left[P\right]^{Tr}
  =  \left[P\right]
\left[\frac{\mu}{b}\right]
\label{P-sym1}
\end{equation}
or
\begin{equation}
\left[P\right] =
\left[\frac{\mu}{b}\right] \left[P\right]^{Tr}\left[\frac{\mu}{b}\right]^{-1},
\label{P-sym2}
\end{equation}
where the diagonal matrix $\displaystyle
\left[\frac{\mu}{b}\right]\equiv\left[\mu\right]\left[b\right]^{-1}$ and
$[\mu]$ is the diagonal matrix of reduced masses $\mu_\Gamma$.

It is clear that the multichannel $P$-matrix is symmetric,
\begin{equation}
\left[P\right] = \left[P\right]^{Tr} ,
\label{P-symmetry}
\end{equation}
if $\displaystyle\left[\frac{\mu}{b}\right]$ is proportional to the unit matrix,
i.e. if the ratio $\displaystyle\frac{\mu_\Gamma}{b_\Gamma}$
is the same in all channels $\Gamma$.

%%%%% Figure captions  %%%%%%%%%%%%
\begin{figure}
\caption{Relative deviation $(b_0-b)/b$ of the natural channel radius
$b_0$ from the exact solution $b$ of eq.~(\protect\ref{frac}) vs $E$
for various values of $N$. Solid lines --- $s$ waves; dash lines ---
$p$ waves; dot-dash lines --- $g$ waves. $\hbar\omega=18$~MeV, reduced
mass $m$ corresponds to the neutron scattering by $A=15$ nucleus.}
\label{fig1}
\end{figure}

\begin{figure}
\caption{$P$-matrix at the natural channel radius for neutron scattered
by $A=15$ nucleus vs energy in the
cases $N=1$ (a) and $N=9$ (b). Solid line --- exact $P$-matrix
calculated by (\protect\ref{Pgen}); dash and dot lines --- discrete
analogues of the $P$-matrix calculated by (\protect\ref{eq:Pdiskr}) and
(\protect\ref{Pdis3}),
respectively; up and down arrows indicate positions of the poles of
the exact $P$-matrix and its discrete analogue,
respectively. Calculations are performed with  $\hbar\omega=18$~MeV
and Woods-Saxon potential of ref.~\protect\cite{Coupl2}.}
\label{fig2}
\end{figure}

\begin{figure}
\caption{$s$-wave phase shift $\delta_0$ of the $p$--$^{15}$N
scattering as a function of the channel radius $b$ used for construction
of the potential $V^{Sh}$ at energies $E=2$~MeV (a) and
$E=10$~MeV (b). The horizontal solid line correspond to the exact value of
$\delta_0$. Phase shifts calculated with $\hbar\omega=18$~MeV and
$N=4$, $N=9$ and $N=19$ are presented by short-dash, solid
and long-dash curves, respectively; dot-dash (dots) curve presents the
results obtained with $\hbar\omega=10$~MeV and $N=9$
($\hbar\omega=26$~MeV and $N=9$). }
 \label{fig3}
\end{figure}

\begin{figure}
\caption{(a) $s$-wave phase shift $\delta_0$ of the $p$--$^{15}$N
scattering as a function of energy $E$. Solid curve shows the
exact values of $\delta_0$. Calculations by the method suggested in this
paper with $\hbar\omega=18$~MeV and channel radius $b=7$~fm,
and $N=10$, $N=8$, $N=6$ and $N=4$ are presented by
the long-dash, dot-dash, dot-dot-dash and short-dash curves,
respectively. Calculations by the method suggested by Kiev
group~\protect\cite{Okhr} are presented by the thin solid curve with
diamonds ($N^{Sh}=50$ and $N=70$) and by dots ($N^{Sh}=8$ and $N=70$).
(b) The same but in the vicinity of the resonance only.}
\label{fig4}
\end{figure}

\begin{figure}
\caption{(a) Wave function in the oscillator representation squared,
$a_{nl}^{2}(k)$, as a function of energy for
$l$=0 and $n=0,1,2$. (b)~The same but in the vicinity of the resonance
only. See fig.~\protect\ref{fig4} for details.}
\label{fig5}
\end{figure}

\begin{figure}
\caption{Radial wave function $\tilde u_l(kr)$
for $s$-wave $p$--$^{15}$N
scattering at energies $E=3$~MeV~(a) and $E=15$~MeV~(b). Solid curve
--- exact; calculations by the method suggested in this
paper with $\hbar\omega=18$~MeV and channel radius $b=7$~fm:
dash curve --- reconstruction with $N=M=10$; dots ---
reconstruction with $N=10$ and $M=100$ [see
eq.~(\protect\ref{reconstruct})].  }
\label{fig6}
\end{figure}

\begin{figure}
\caption{$p$-wave phase shifts of the $p$--$^{15}$N
scattering~(a) and the corresponding wave functions in the oscillator
representation squared, $a_{nl}^{2}(k)$, for
$l$=1 and $n=1$~(b), as functions of energy.
See fig.~\protect\ref{fig4} for details.}
\label{fig7}
\end{figure}

\begin{figure}
\caption{The same as fig.~\protect\ref{fig7} but for $d$-waves.
\protect\hphantom{The same as fig.~\protect\ref{fig7} but for
$d$-waves. $d$-waves.}}
\label{fig8}
\end{figure}

%\end{document}

%%%%% Figures %%%%%%%%
\newpage
\begin{figure}
\psfig{file=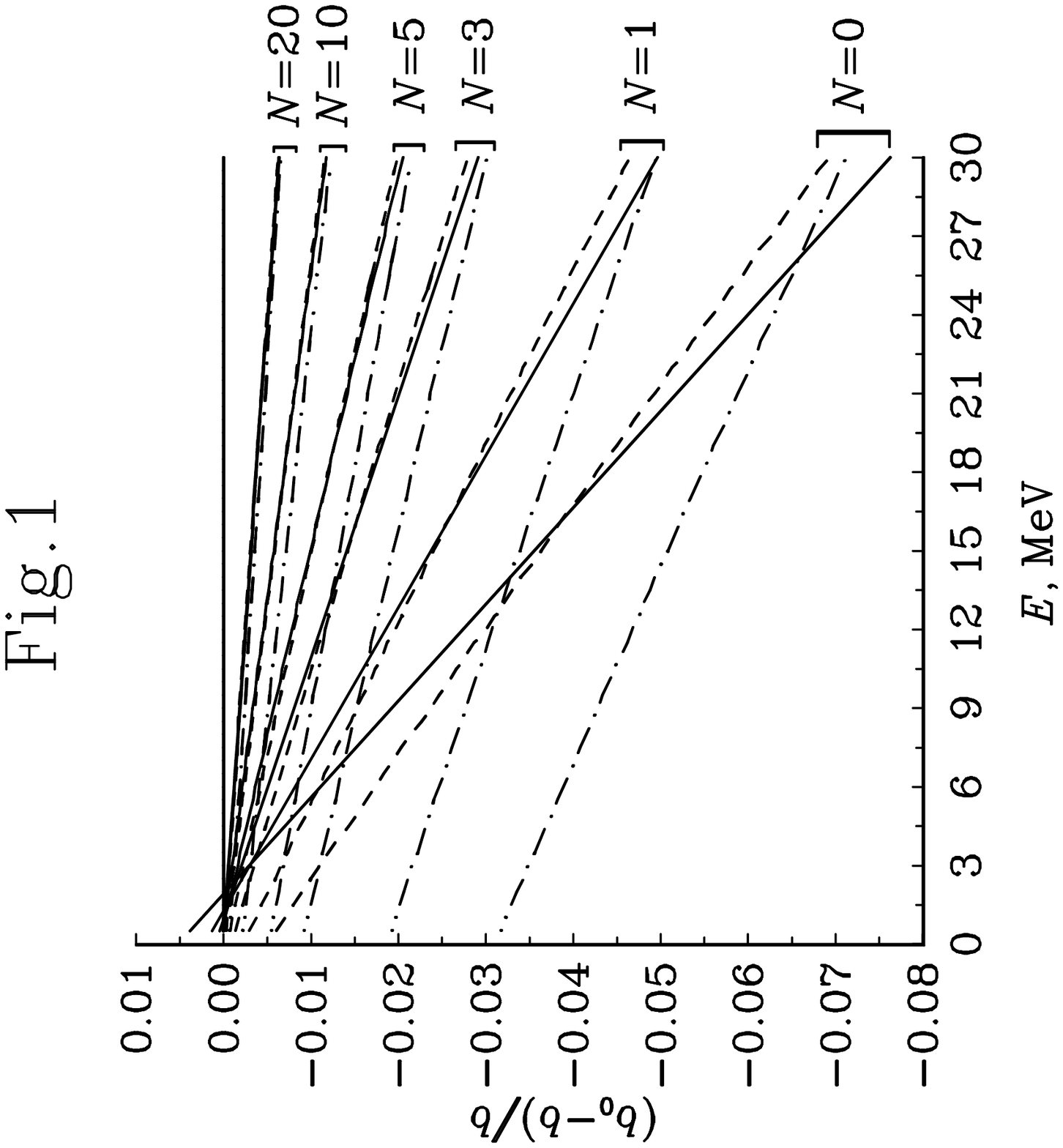}
%\vspace{5cm}
\end{figure}

\newpage
\begin{figure}
\psfig{file=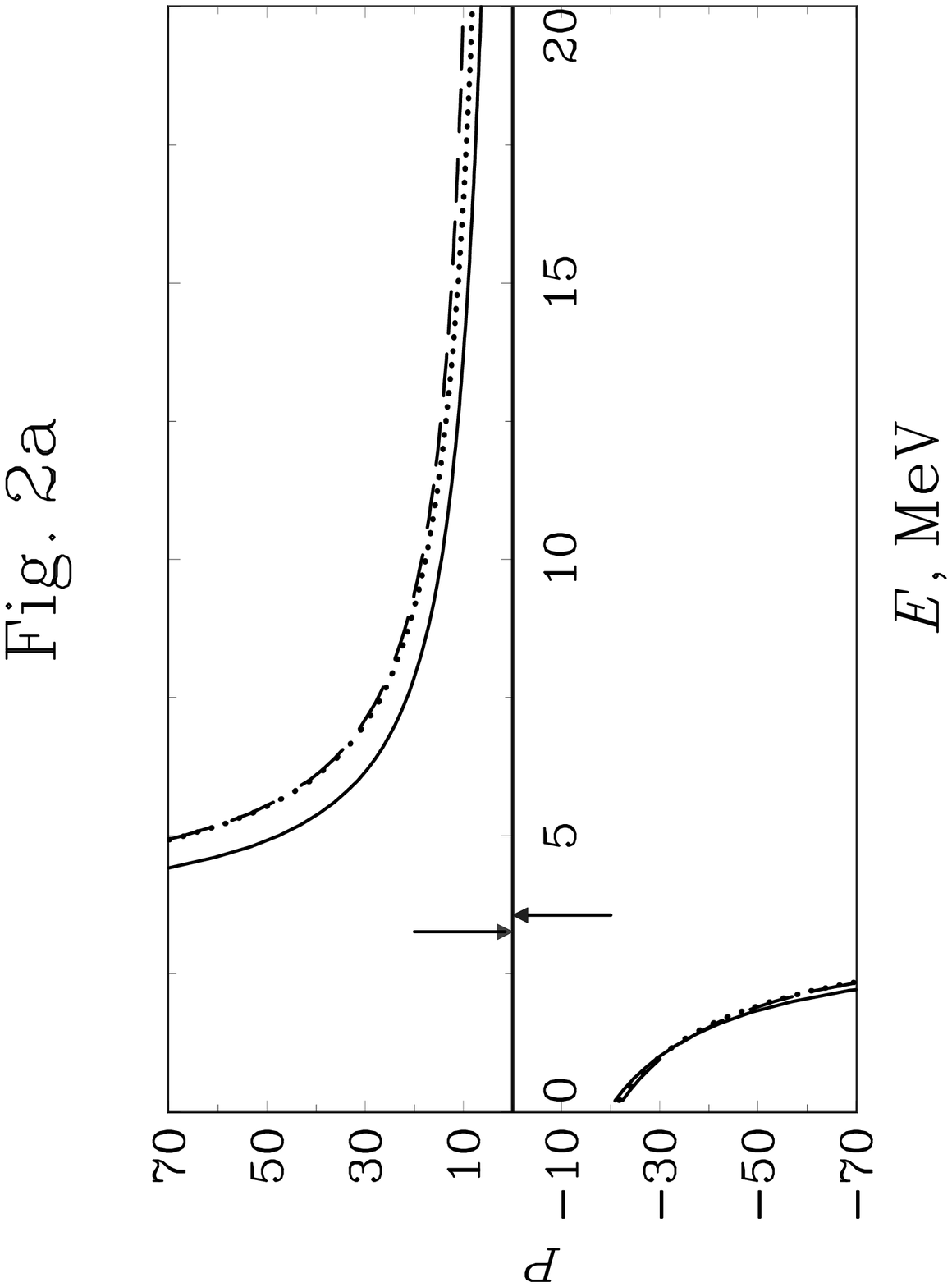}
%\vspace{5cm}
\end{figure}

\newpage
\begin{figure}
\psfig{file=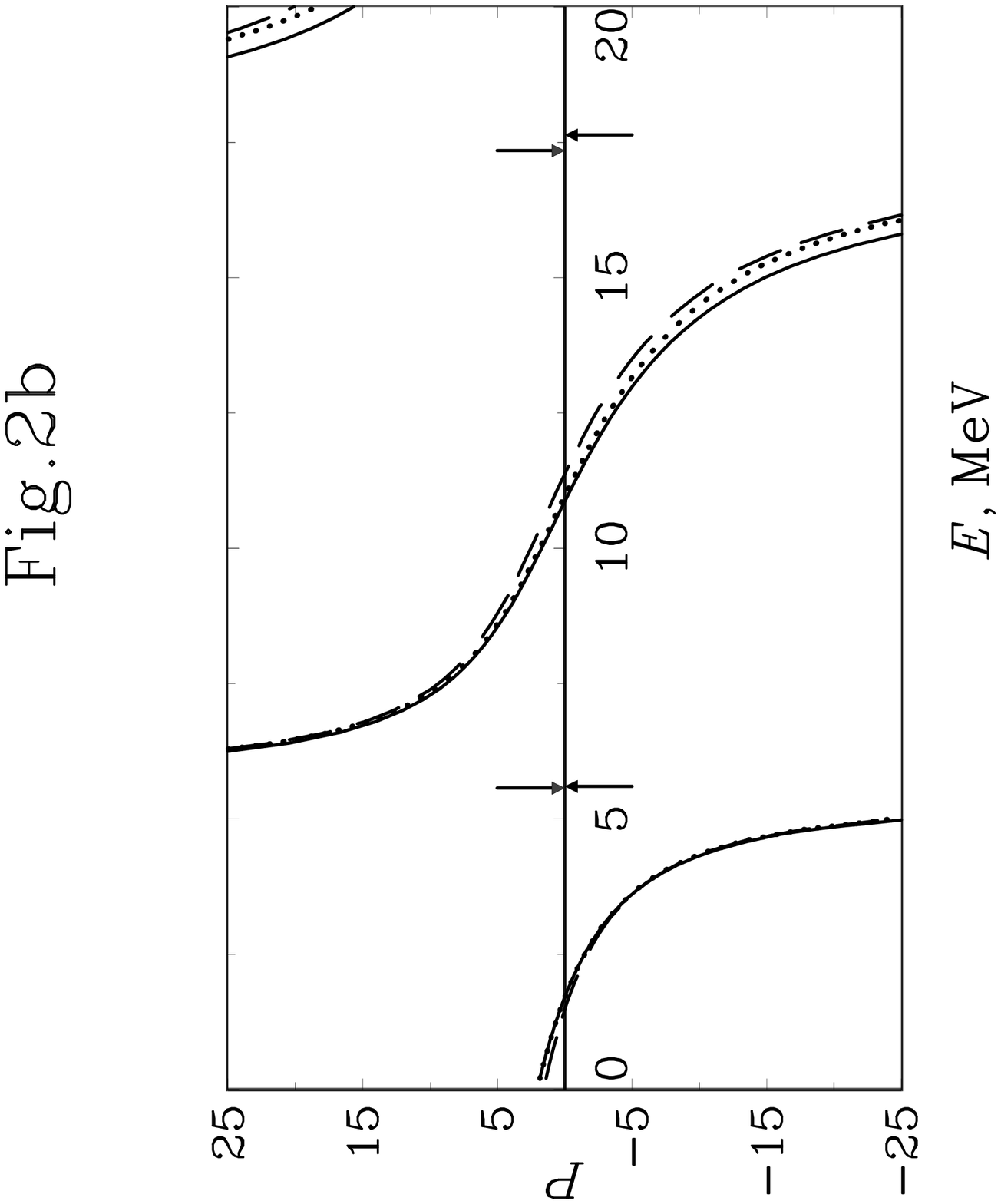}
%\vspace{5cm}
\end{figure}

\newpage
\begin{figure}
\psfig{file=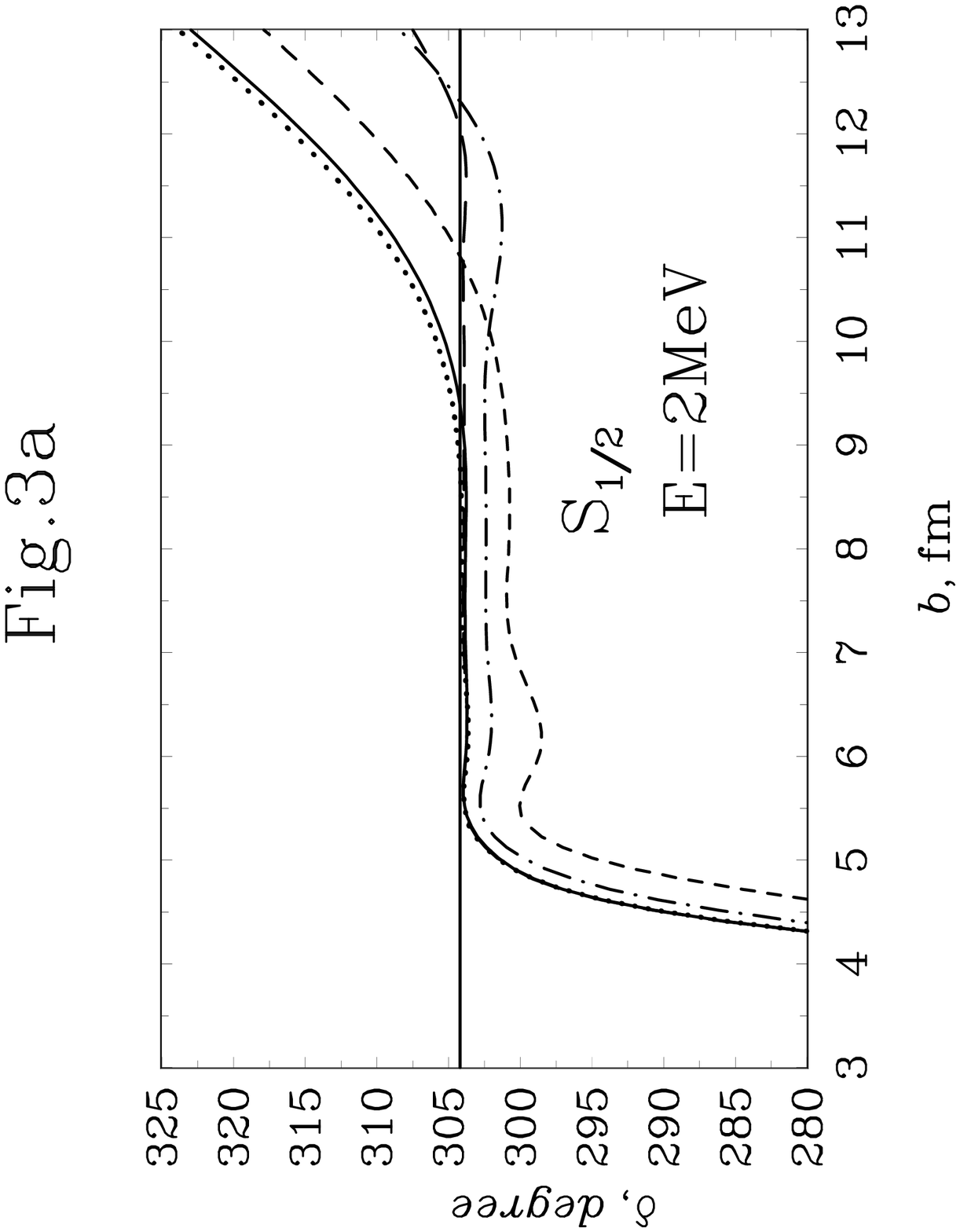}
%\vspace{5cm}
\end{figure}

\newpage
\begin{figure}
\psfig{file=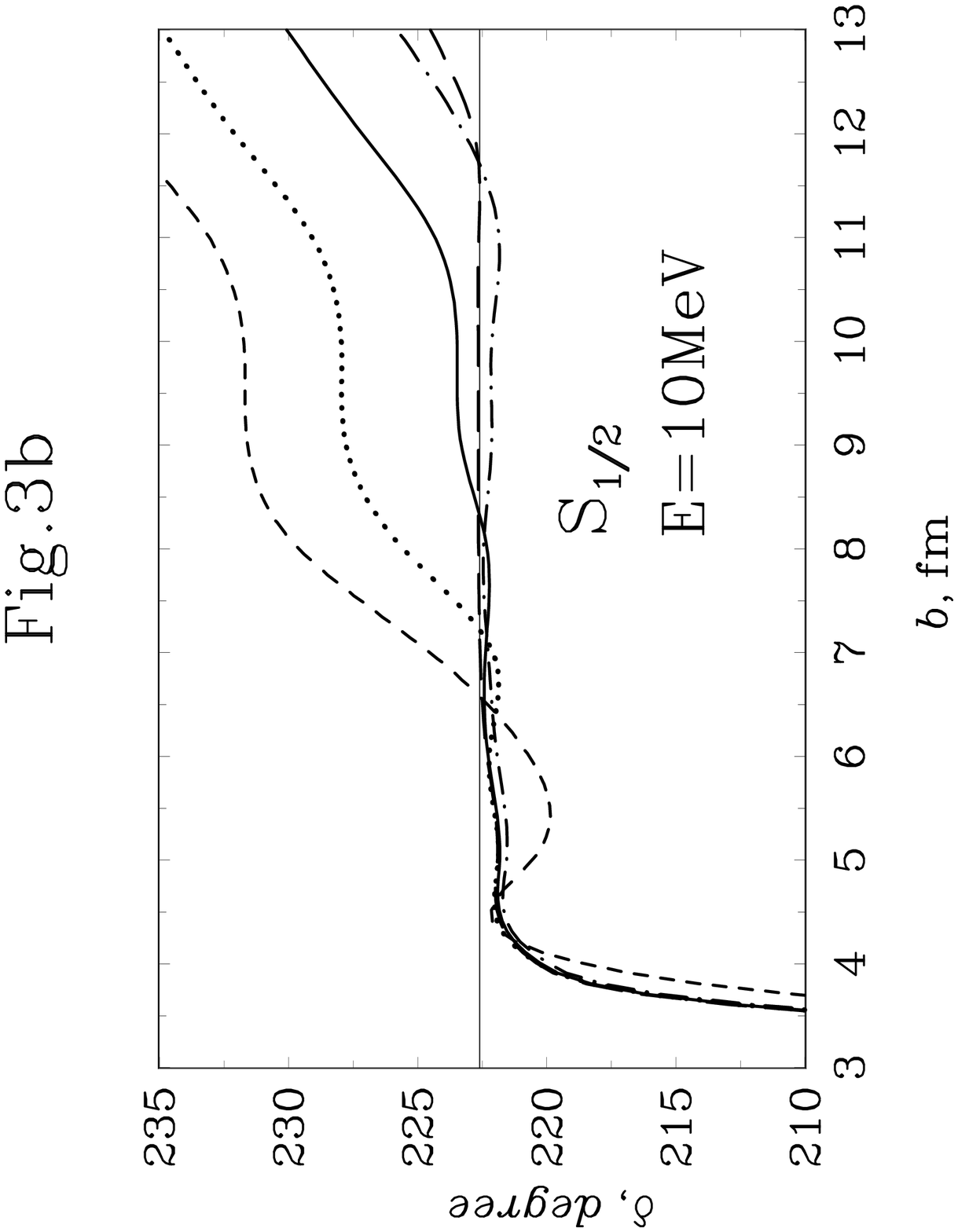}
%\vspace{5cm}
\end{figure}

\newpage
\begin{figure}
\psfig{file=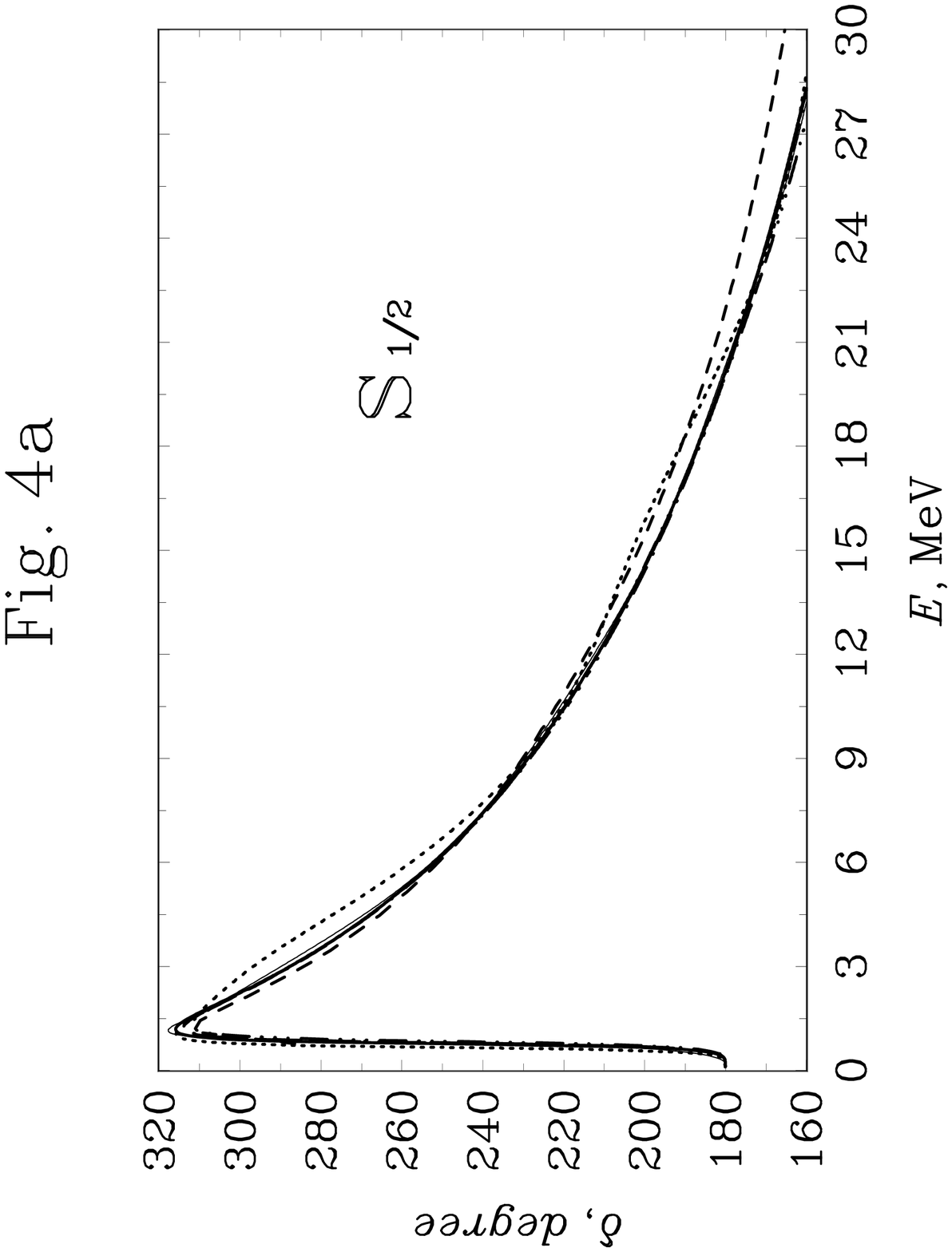}
%\vspace{5cm}
\end{figure}

\newpage
\begin{figure}
\psfig{file=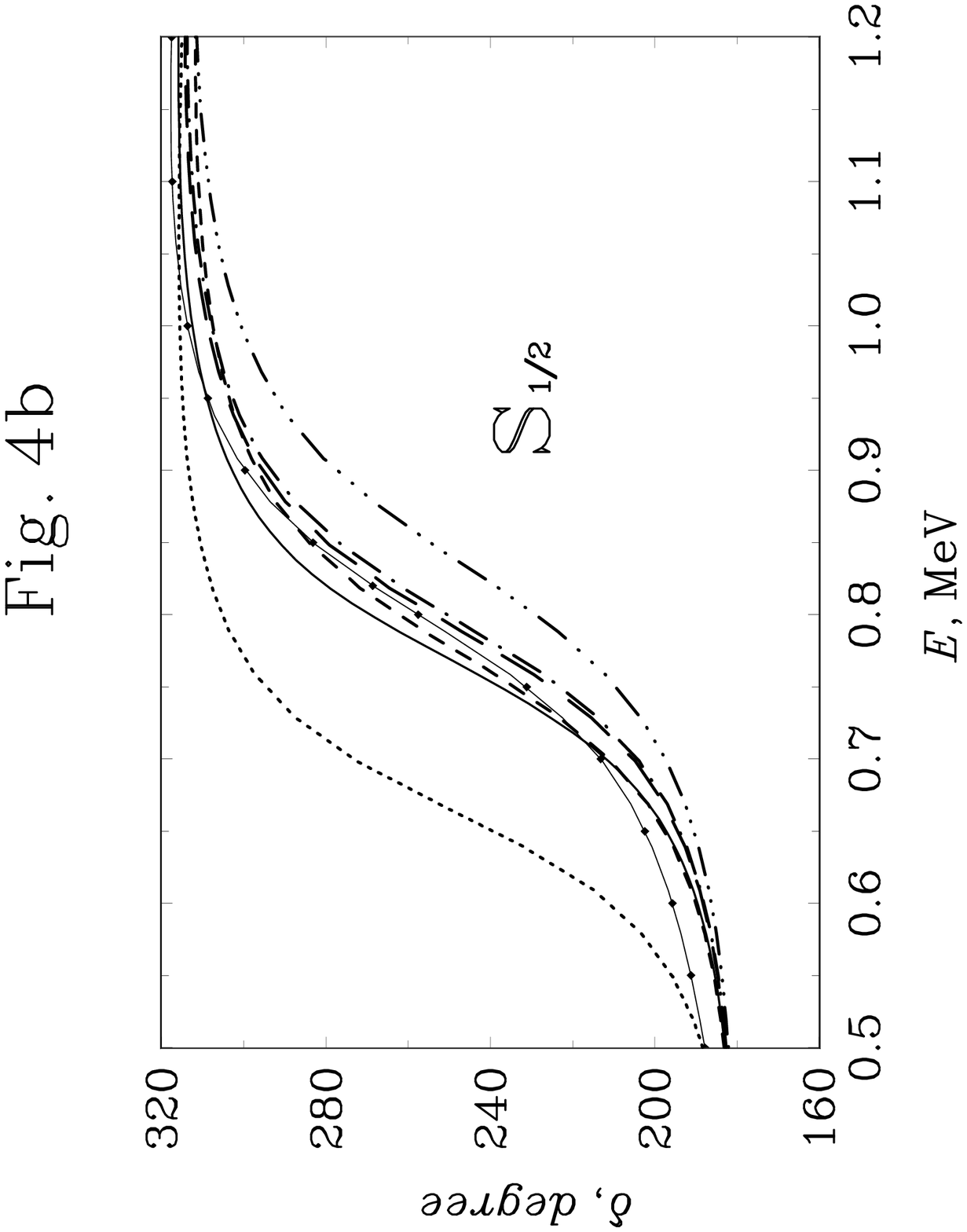}
%\vspace{5cm}
\end{figure}

\newpage
\begin{figure}
\psfig{file=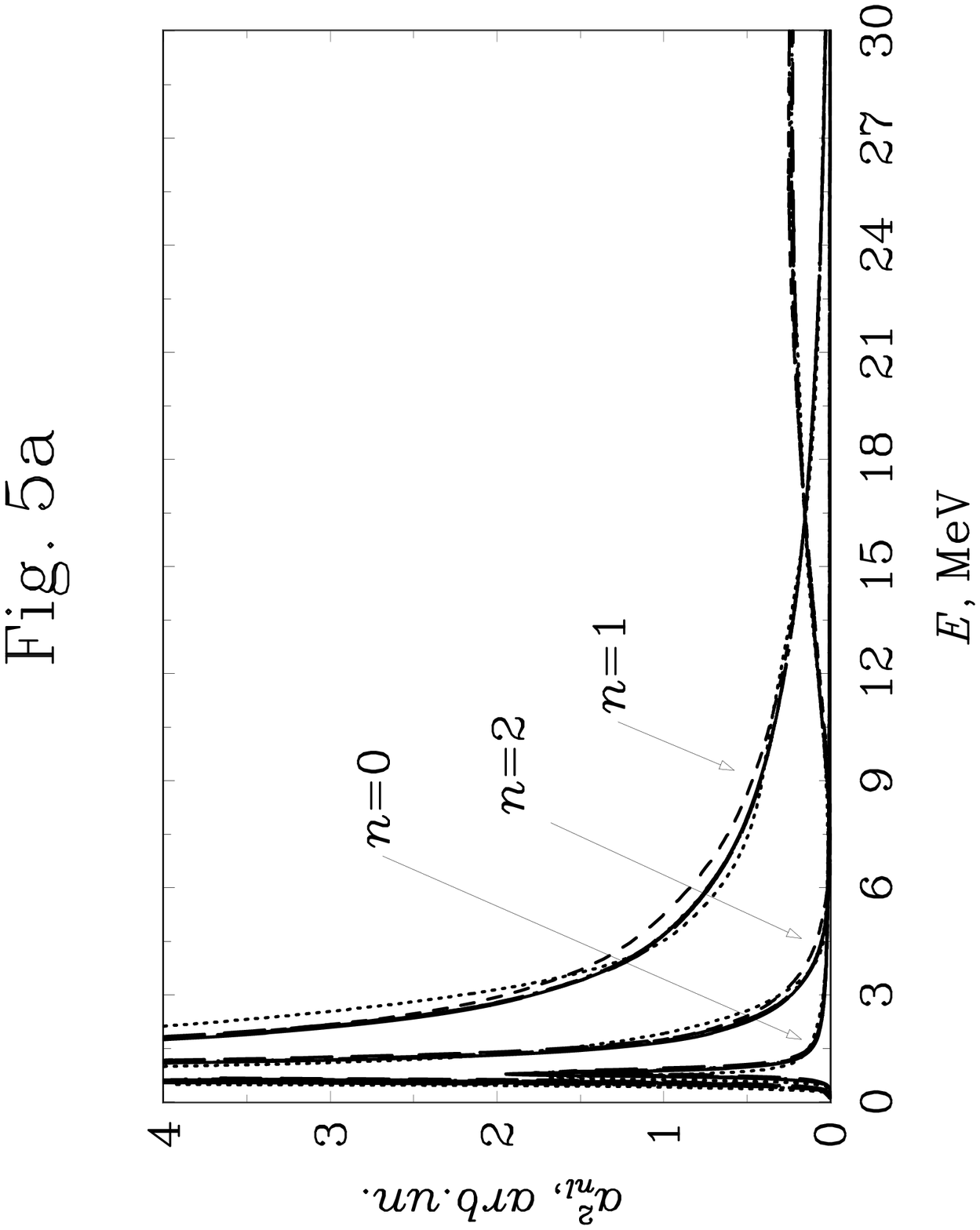}
%\vspace{5cm}
\end{figure}

\newpage
\begin{figure}
\psfig{file=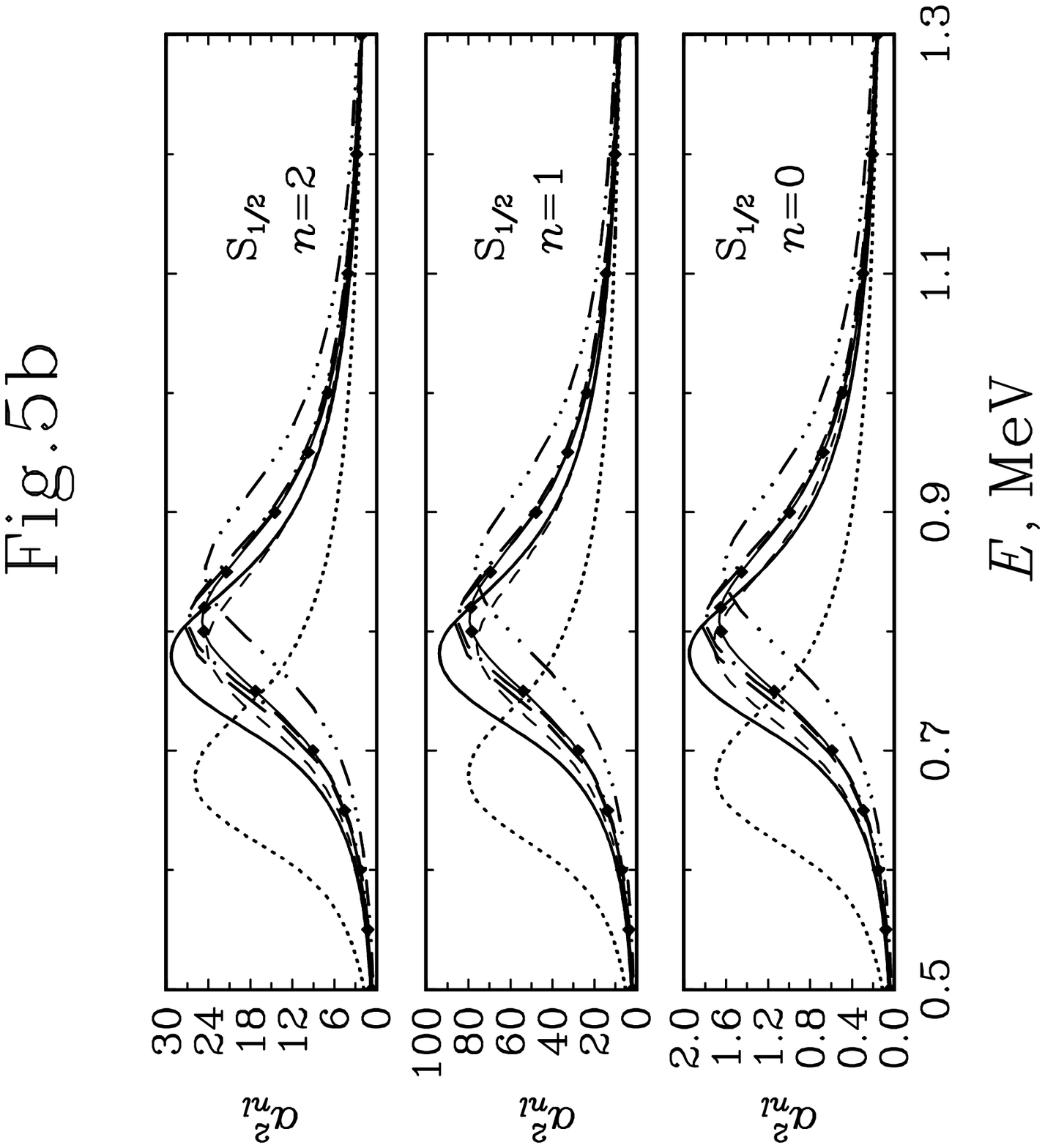}
%\vspace{5cm}
\end{figure}

\newpage
\begin{figure}
\psfig{file=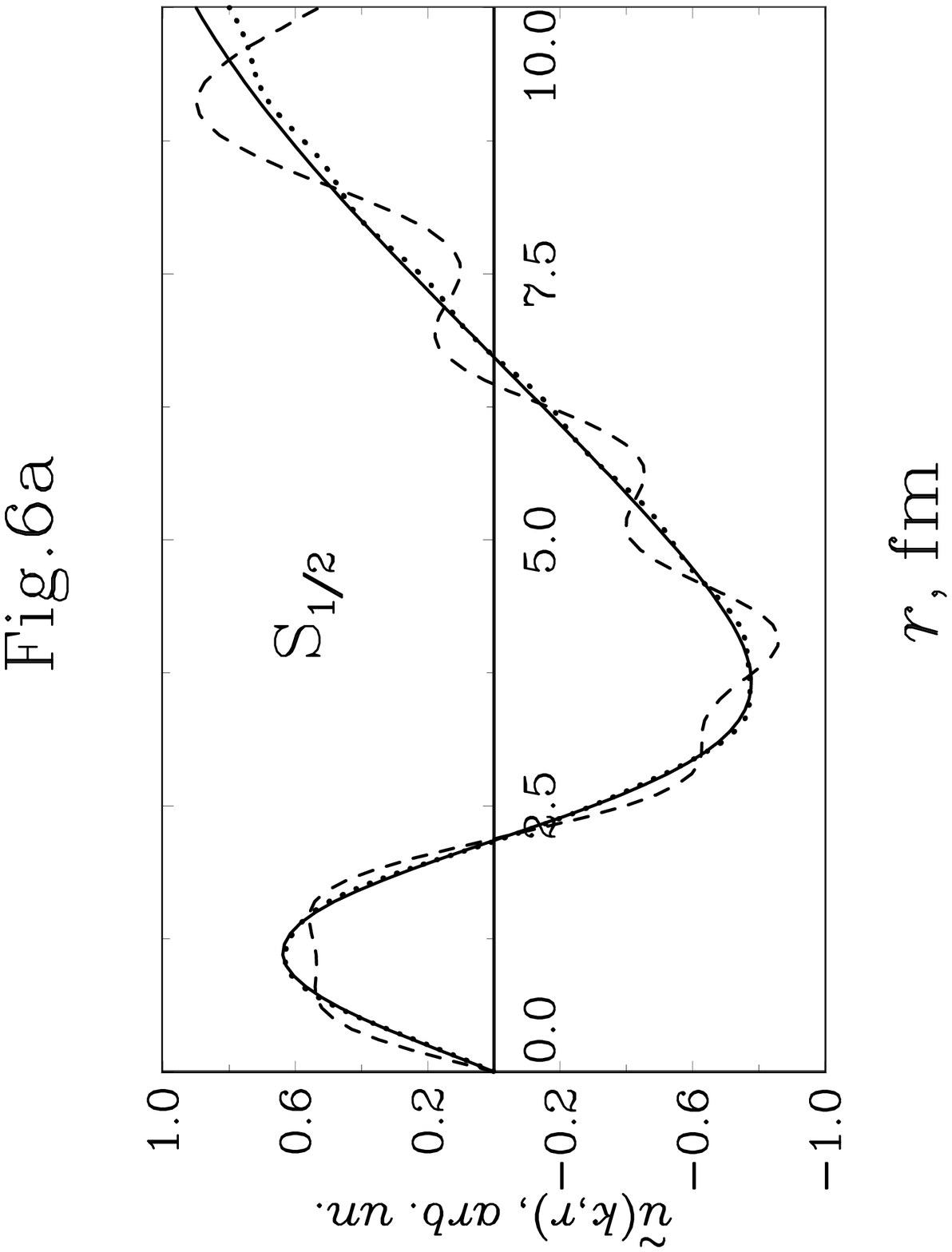}
%\vspace{5cm}
\end{figure}

\newpage
\begin{figure}
\psfig{file=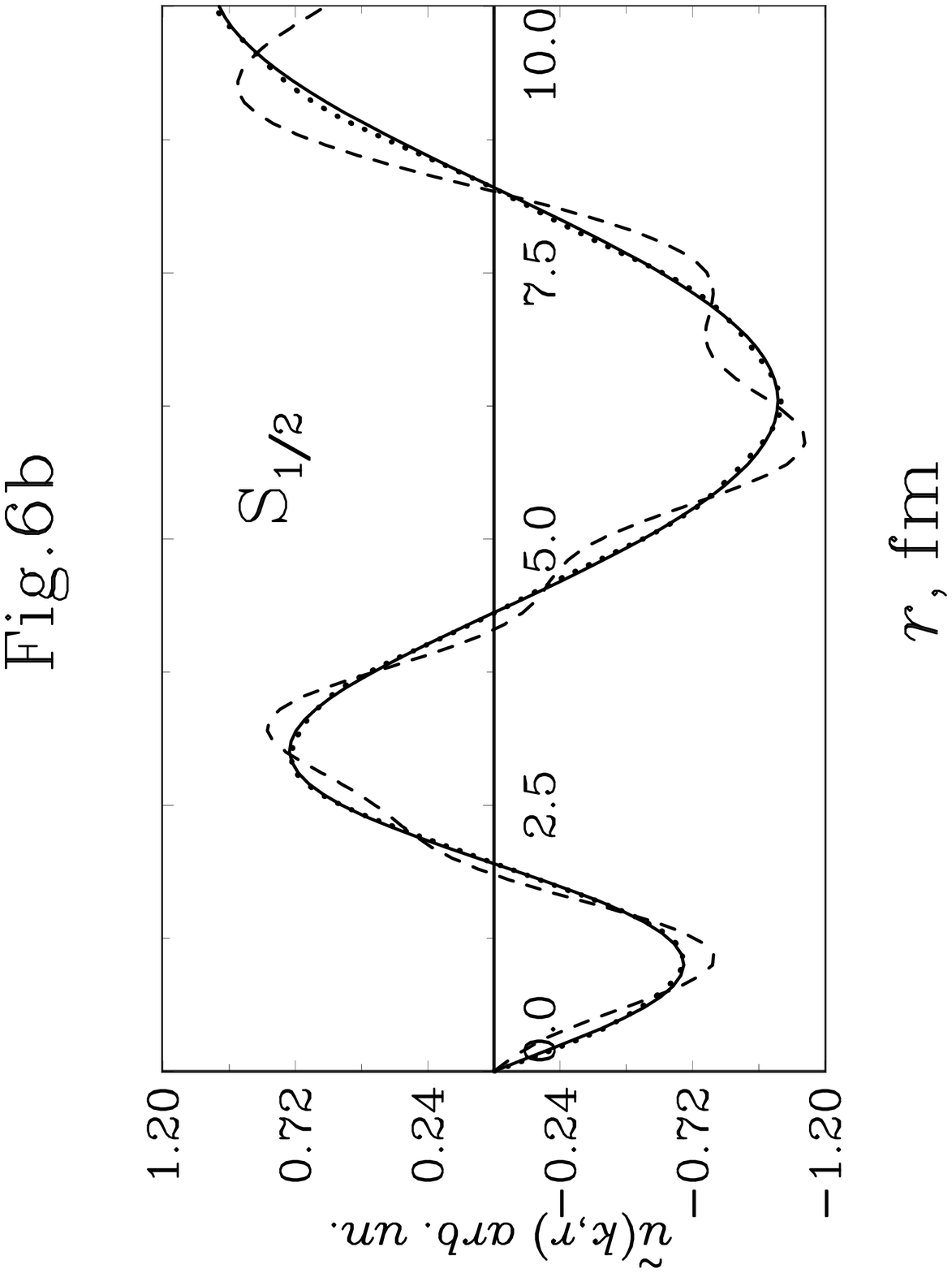}
%\vspace{5cm}
\end{figure}

\newpage
\begin{figure}
\psfig{file=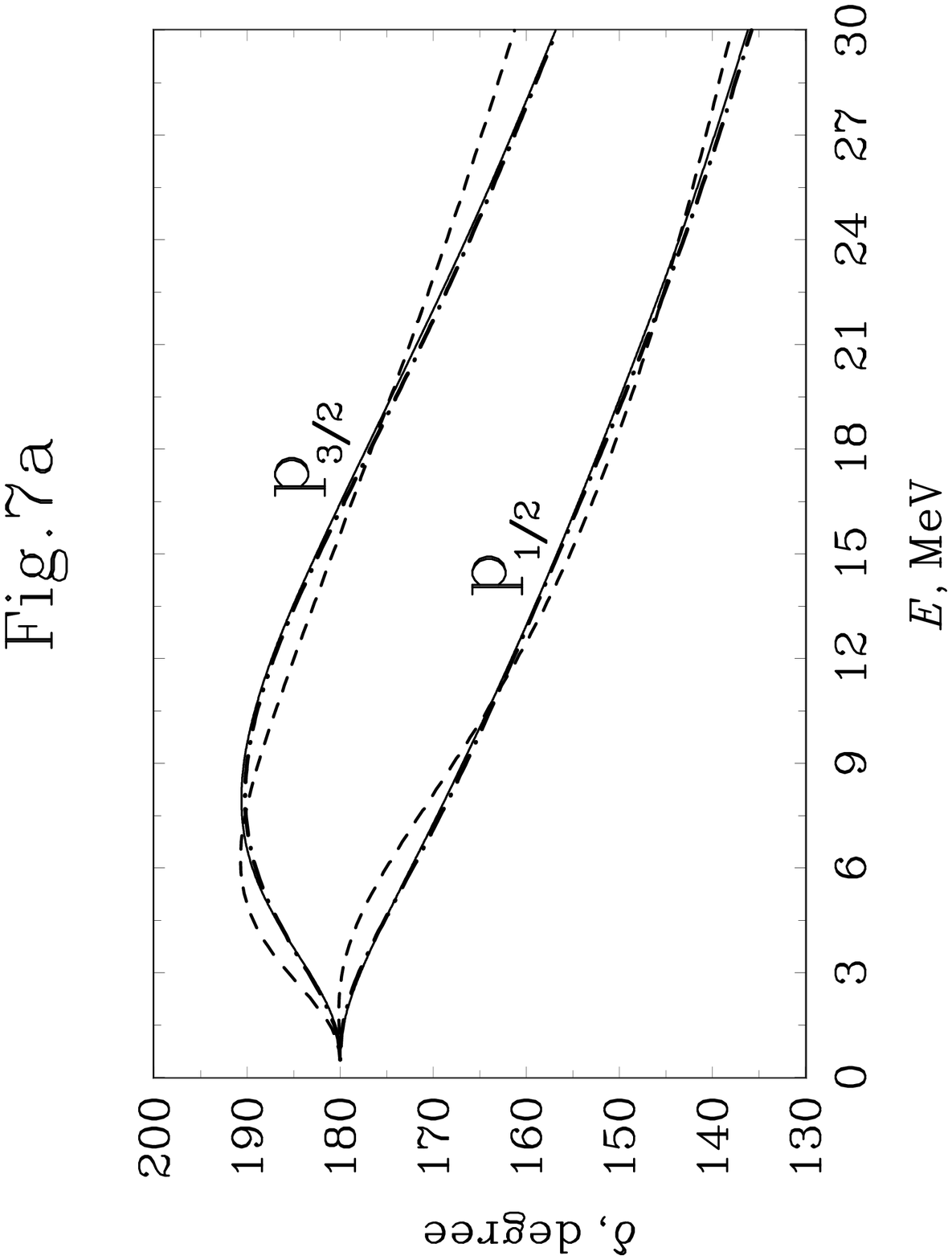}
%\vspace{5cm}
\end{figure}

\newpage
\begin{figure}
\psfig{file=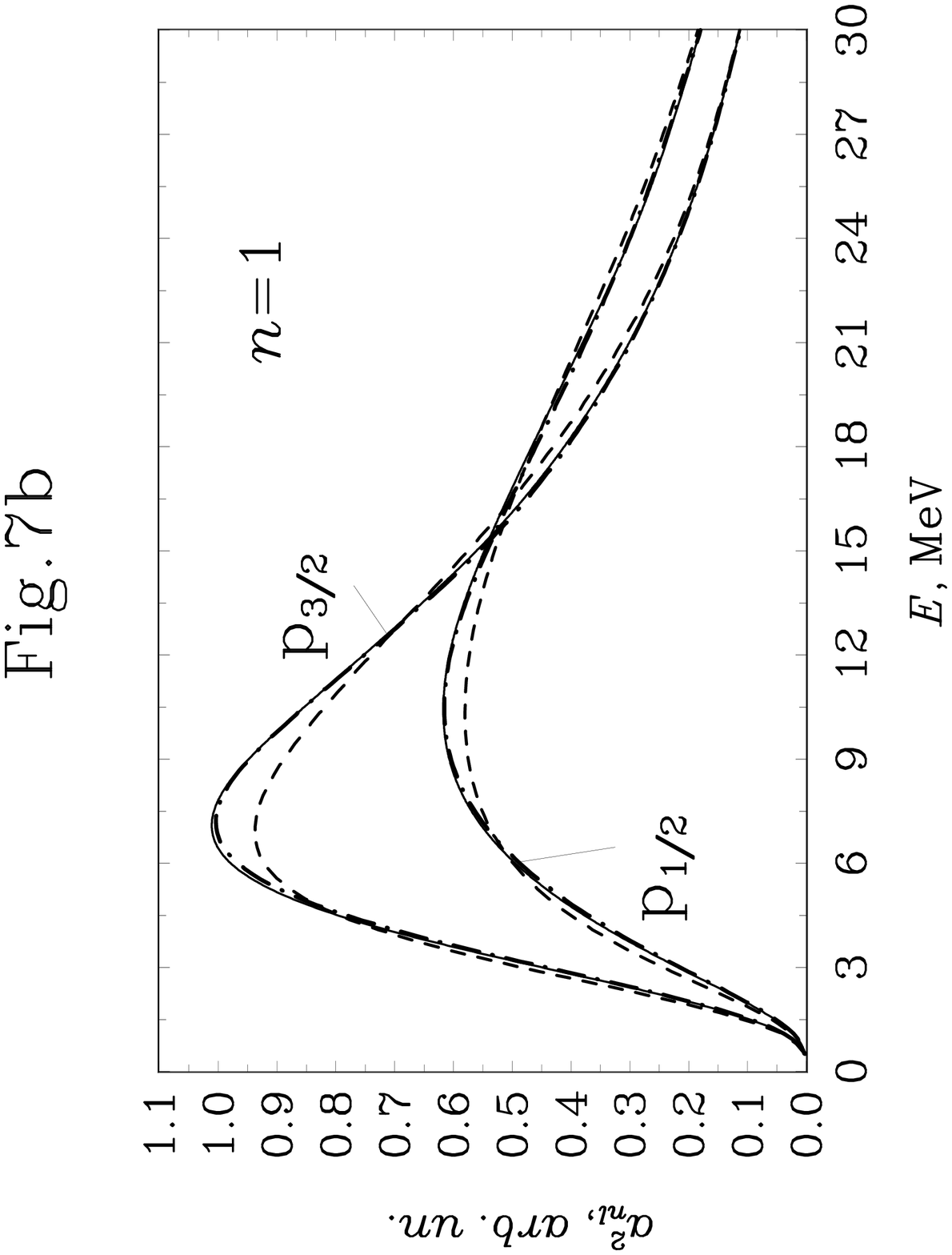}
%\vspace{5cm}
\end{figure}

\newpage
\begin{figure}
\psfig{file=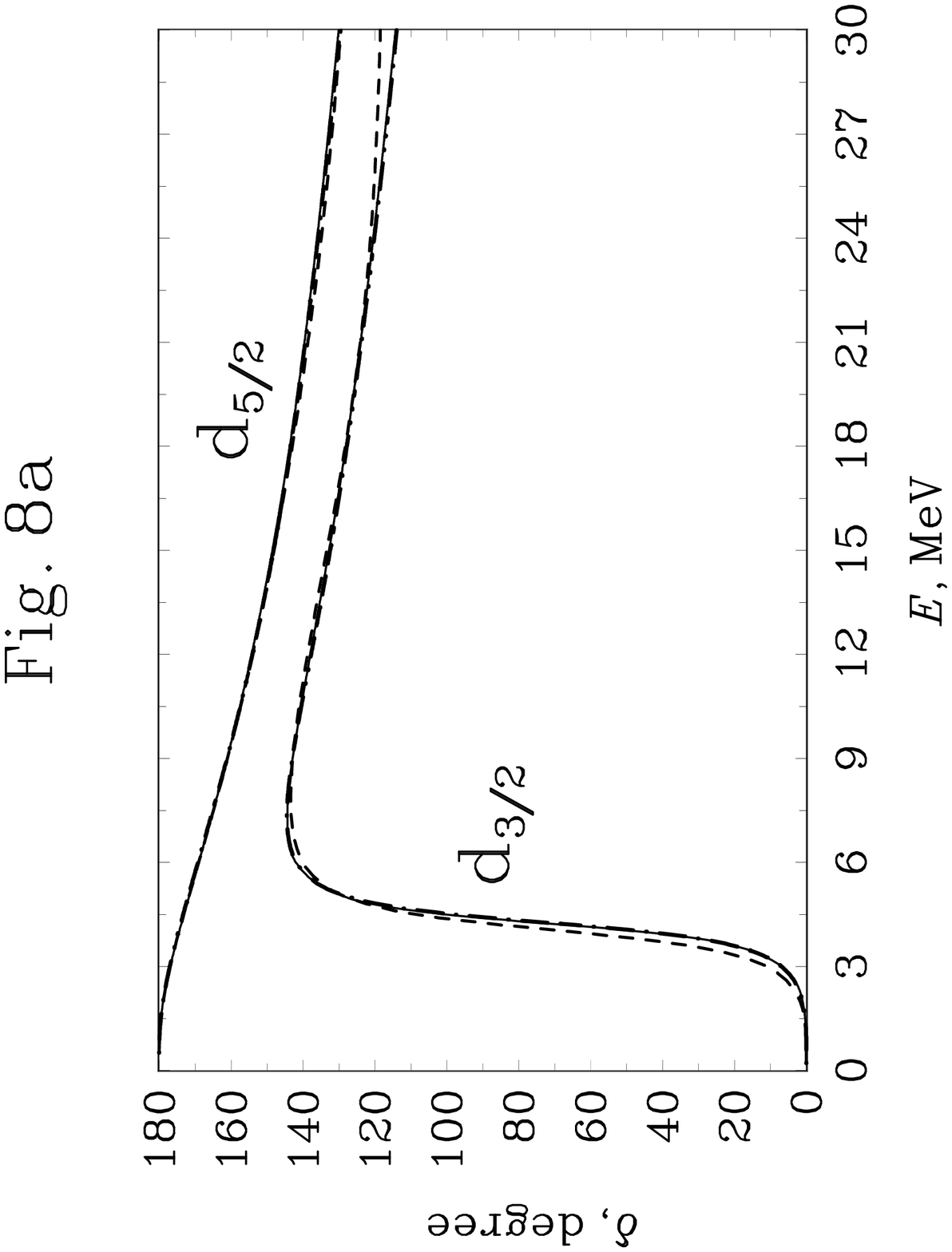}
%\vspace{5cm}
\end{figure}

\newpage
\begin{figure}
\psfig{file=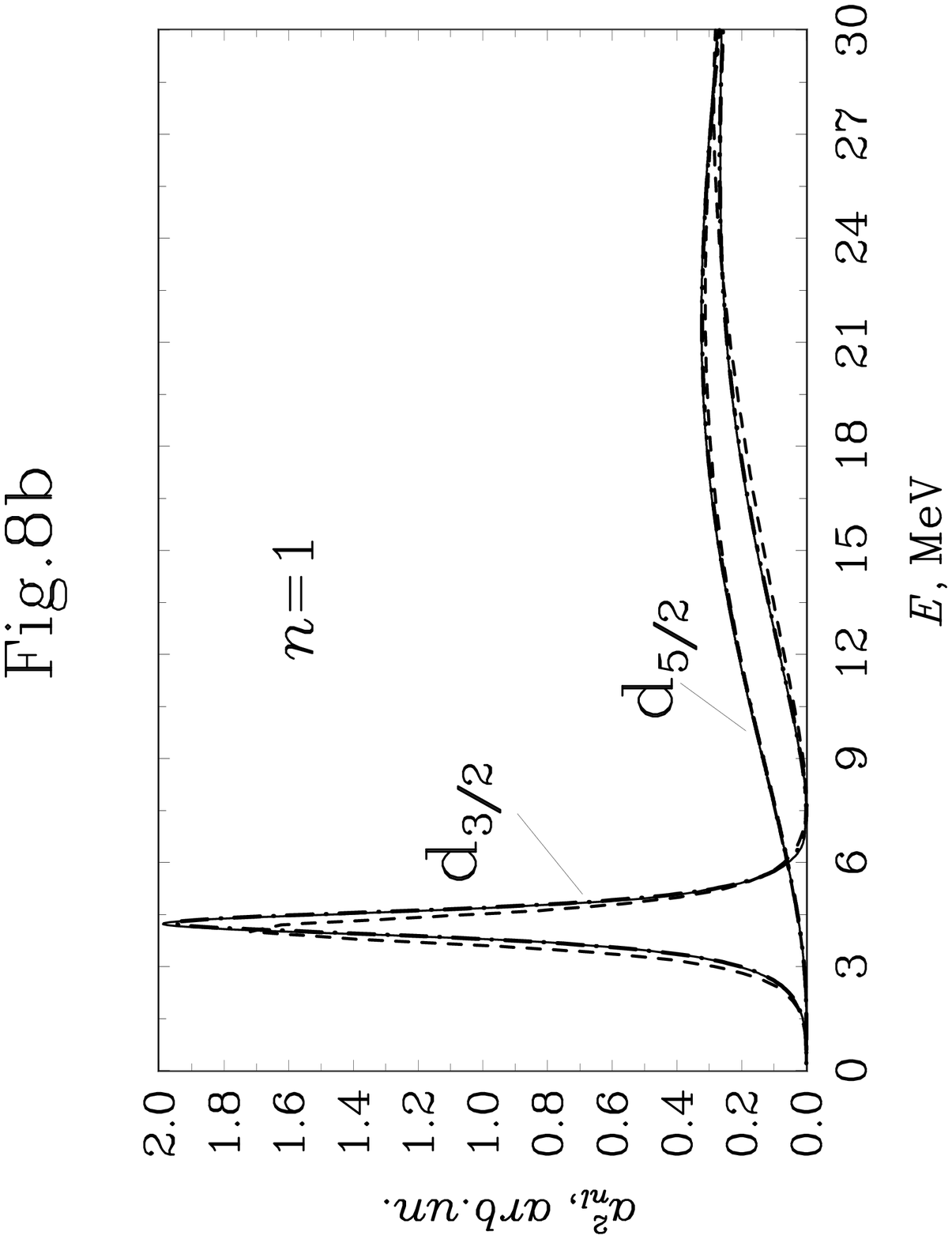}
%\vspace{5cm}
\end{figure}

\end{document}